\documentclass[english,aip,manuscript=article]{revtex4-1}

\usepackage{graphicx}
\usepackage{bm}
\usepackage{amsmath}
\usepackage[utf8]{inputenc}
\usepackage[T1]{fontenc}
\usepackage{mathptmx}
\usepackage{physics}
\usepackage{float}
\usepackage{hyperref}
\usepackage{xcolor}
\usepackage{soul}
\usepackage{comment}
\usepackage{array}
\usepackage{booktabs}
\makeatletter
\newcounter{subsubsubsection}[subsubsection]

\newcommand\subsubsubsection{\@startsection{subsubsubsection}{4}{\z@}%
  {-3.25ex\@plus -1ex \@minus -.2ex}%
  {1.5ex \@plus .2ex}%
  {\normalfont\normalsize\itshape}}
\newcommand\l@subsubsubsection{\@dottedtocline{4}{7em}{4em}}

\newcommand{\bR}{{\bm R}}
\newcommand{\br}{{\bm r}}
\newcommand{\bP}{{\bm P}}

\newcommand{\bp}{{\bm p}}

\newcommand{\bG}{{\bm G}}
\newcommand{\bX}{{\bm R}}
\newcommand{\bDelta}{{\bm \Delta}}

\newcommand{\bGamma}{{\bm \Gamma}}

\newcommand{\hH}{\hat H}

\newcommand{\hbr}{{\hat{\bm r}}}

\newcommand{\bB}{{\bm B}}

\newcommand{\pp}[2]{\frac{\partial {#1}}{\partial {#2}}}
\newcommand{\hbm}[1]{\hat{\bm{#1}}}
\makeatother

\begin{document}
\author{Zhen Tao}
\email{zhen.tao@uri.edu}
\affiliation{Department of Chemistry, University of Rhode Island, Kingston, Rhode Island, 02881, USA}
\author{Mansi Bhati}
\affiliation{Department of Chemistry, Princeton University, Princeton, New Jersey 08544, USA}
\author{Joseph E. Subotnik}
\email{js8441@princeton.edu}
\affiliation{Department of Chemistry, Princeton University, Princeton, New Jersey 08544, USA}

\title{ Non-Resonant Raman Optical Activity From Phase-Space Electronic Structure Theory} 

\begin{abstract}
   In order to model experimental non-resonant Raman optical activity, chemists must compute a host of second-order response tensors, (e.g. the electric-dipole magnetic-dipole polarizability) and their nuclear derivatives along  a set of vibrational modes.   While these response functions are almost always computed  within
   a Born-Oppenheimer (BO) framework, here we provide a natural interpretation of the electric-dipole magnetic-dipole polarizability  within  phase space electronic structure theory, a beyond-BO model whereby the electronic structure depends on nuclear momentum ($\bP$) in addition to nuclear position ($\bR$). By coupling to nuclear momentum, phase space electronic structure theory is able to capture the asymmetric response of the electronic  properties to an external field, in sofar as  {\em for a vibrating (non-stationary) molecule},
   $\frac{\partial \bm{\mu}}{\partial  \bm{B}} \ne 
   \frac{\partial \bm{m}}{\partial  \bm{F}}$  
where $\bm{\mu}$ and $\bm{m}$ are the electrical linear and magnetic dipoles, and $\bm{F}$ and $\bm{B}$ are electric and magnetic fields. As an example, for a prototypical methyloxirane molecule,  we show that phase space electronic structure theory is able to deliver a reasonably good match with experimental results in a manner that is invariant to gauge origin $\bG_0$. 
\end{abstract}

\maketitle

\section{Introduction}
Chirality has important consequences with regards to biological protein recognition, and the capacity to influence the chirality forms the basis of major fields of research, including  asymmetric catalysis for enantioselective synthesis,\cite{NobelList,NobelMacMillan,bolm_introduction_2003} chiral optoelectronics,\cite{long_chiral-perovskite_2020,crassous_materials_2023} and biosensing for drug-target interactions.\cite{liu_soft_2020,warning_nanophotonic_2021} Chirality has been more recently discovered to couple with electronic spins,\cite{naaman_chiral-induced_2012,bloom_chemical_2024} termed chiral-induced spin selectivity (CISS), which has important implications in spintronics,\cite{naaman_spintronics_2015} electrocatalysis,\cite{naaman_chiral_2020,sang_chirality_2022} and possibly the origin of homochirality in life.\cite{ozturk_origins_2022,ozturk_chirality-induced_2023}

At the most basic experimental level, chiral isomers can be distinguished by their difference in optical responses, so-called  chiroptical spectroscopies. Circular dichroism (CD) is arguably the most common chiroptical technique, whereby one measures the difference in absorption of left-circularly polarized (LCP) light and right-circular polarized (RCP) light, and CD is usually studied in two distinct forms: electronic CD (ECD) which uses ultraviolet (UV) light to probe the electronic transitions and vibrational CD (VCD) which uses infrared (IR) light to probe the vibrational transitions. Apart from CD,  Raman optical activity (ROA) is another means to measure chirality, whereby one  measures the difference in Raman-scattered intensity of LCP and RCP light. Unlike VCD, ROA is particularly effective for samples in aqueous solutions because neat water produces weak Raman signals, making ROA a powerful technique to study biomolecules such as proteins, carbohydrates, and nuclei acids in their natural aqueous solutions.\cite{barron_cosmic_2012} In practice, because Raman signals are weak, experimentalists often boosts ROA signals by selectively coupling the  electronic transition to vibrations and surface-enhanced Raman scattering.\cite{nie_probing_1997,kneipp_single_1997,abdali_surface_2008}

In this article, our focus will be on ROA (as opposed to VCD). That being said,  as  will be made clear below, the motivation for the present article is to transfer advanced techniques from  the theory of VCD over to the realm of ROA (see below). In that regard, a few preliminary words about VCD are especially pertinent at this juncture. 
(Interestingly, because the VCD and ROA signals are so weak, the theory behind both of these chiroptical spectroscopies were developed at roughly the same time as the experimental apparatuses themselves in the 1970s.\cite{barron_rayleigh_1971,barron_raman_1973,hug_raman_1975})

As a brief reminder, within VCD, when averaged over all incoming directions for the light, 
the rotational strength   is $\mathcal{R}=\hat{\bm \mu}^{tot}_{i\rightarrow f}\cdot\hat{\bm m}^{tot}_{i\rightarrow f}$,\cite{Nafie1983,nafie_infrared_1997}. Here, $\hat{\bm \mu}^{tot}_{i\rightarrow f}$ is  the total, linear transition dipole moments (include both electronic and nuclear contributions); $\hat{\bm m}^{tot}_{i\rightarrow f}$ is  the total magnetic transition dipole moments (include both electronic and nuclear contributions).
Note that,  within standard quantum chemistry dogma, all vibrational spectroscopy (including VCD) is  described in terms of matrix elements between vibrationally excited states on a given Born-Oppenheimer (BO) electronic surface.
Within such a BO approximation, all dynamic electron-nuclear correlation is ignored and the electronic component of the magnetic transition dipole vanishes, $\hat{\bm m}^{e}_{i\rightarrow f} = 0$.  As a result,  the computed  VCD signals is proportional to the the magnetic moment from bare nuclear charges vibrating -- which is completely wrong. In other words, a reasonable model of VCD spectroscopy mandates that one account for dynamics on more than one BO surface.  For going on  three decades, the common solution to this is known as magnetic field perturbation theory,\cite{Stephens_1985_MFP,stephens_gauge_1987,stephens_determination_2000} which computes the necessary 
electronic atomic axial tensor (AAT) through a Berry curvature expression (see Eq. \ref{eq:AAT} below).\cite{Mead1979,Berry1984,resta_molecular_2023}  Here, the Berry curvature accounts for the additional electronic screening to the nuclear transition magnetic dipole moment and can be computed using several different  BO calculations as input.

Now, in a recent paper,\cite{duston_phase-space_2024} our research group  demonstrated an alternative approach to  VCD spectroscopy, whereby one can  account for non-BO effects without computing the Berry curvature. Our approach  was to completely avoid the BO framework and  
instead diagonalize an electronic phase space (PS) Hamiltonian\cite{tao_basis-free_2025} that captures the local dynamic electron–nuclear coupling via ${\bm P}\cdot \hat{\bm \Gamma}$ term; here,
${\bm P}$ is the nuclear momentum and $\hat{\bm \Gamma}$
is an electronic operator that captures the local electronic momentum. For the form of the proposed  $\hat{\bm \Gamma}$ operator, see below (Eqs. \ref{eq:gammasum_mag} - \ref{eq:erf_mag}); for a meaningful operator, $\hat{\bm \Gamma}$   must satisfy the same four symmetry constraints as does the exact electron-nuclear derivative coupling operator.\cite{Wu2022} See  Eqs. \ref{eq:dconstrain1a}-\ref{eq:dconstrain4a} and Eqs. \ref{eq:Gamma1}-\ref{eq:Gamma4} below for these constraints. In practice, we have made progress by building a $\hat{\bm \Gamma}$ operator that partitions space and lets electrons couple to  nearby nuclear motion.  As shown in Ref. \citenum{tao_basis-free_2025}, by working with a phase space Hamiltonian  -- that is parametrized by both $\bm{R}$ and $\bm{P}$ -- we can  naturally recover the non-zero electronic momentum dragged by nuclear motion. We can also recover reasonable electronic current densities and VCD signals.\cite{duston_phase-space_2024,tao_electronic_2024} 
Lastly, all of the theory above can be and has been extended to 
systems in the presence of an external magnetic field,\cite{bhati_phase-space_2025-1,bhati_phase-space_2025} which introduces new physics as well.

At this point, let us return to ROA. According to standard  time-dependent perturbation theory, the three key optical activity tensors that dictate ROA signals are the electric-dipole electric-dipole polarizability $\bm \alpha$,  the electric-dipole magnetic-dipole polarizability $\bm G$, and the electric-dipole electric quadrupole polarizability $\bm A$. For practical computations of nonresonant ROA, the theory of ROA has historically made several approximations:
\begin{enumerate}
    \item The molecular response to the electromagnetic fields is assumed to be linear, and one can calculate the  relevant response function with BO theory through nuclear derivatives of the Berry curvature.\cite{amos_electric_1982} 
    \item Nuclear vibrations are assumed to be harmonic.
    \item In the static limit, the ROA tensors are assumed to be independent of the frequency of the incident light.
    \item The ROA tensors are expanded to the first order in displacement with respect to the normal modes coordinates within the Placzek approximation\cite{placzek_rayleigh_1959}.
    \item Within the perturbation theory, many-body BO states are simplified to single-particle states.
\end{enumerate}

One can go beyond these assumptions, including anharmonicity,\cite{cornaton_analytic_2016,luber_raman_2017,yang_anharmonic_2022} many-body effects through the more accurate treatment of electron-electron correlation,\cite{rappoport_lagrangian_2007,crawford_coupled-cluster_2011} and solvent effects,\cite{hopmann_explicit_2011,tmutter_calculation_2015} but it must be noted that the theory of ROA is already quite difficult to implement and expensive to calculate (even more expensive than for VCD).\cite{ruud_theoretical_2009,cheeseman_basis_2011}

With this background in mind, the motivation of the present article is clear. Our aim is to apply the phase space methods used in Ref. \citenum{duston_phase-space_2024} for VCD theory to the realm of ROA spectroscopy. Here we limit our discussion in the vibrational non-resonant ROA and we show that the imaginary component of the electric-dipole magnetic-dipole polarizability $\bm G'$ tensor can be expressed naturally within a phase space perspective, a notion that was hinted at previously by   Escribano, Freedman, and Nafie \cite{escribano_bond_1987}. Specifically, we will show that one can directly determine $\bm G'$ through nuclear momentum dependence, which corresponds to the asymmetric response of the electronic transition electric dipole moment ($\bm \mu$
) and magnetic dipole moment ($\bm m$) to the external magnetic field and electric field, respectively, for a vibrating molecule.
An outline of this article is as follows: In Sec. \ref{sec:BO}, we review the traditional way of computing the ROA tensor $\bm G'$.
Thereafter, in Sec. \ref{sec:PS}, we derive the PS framework to evaluate the ROA tensor $\bm G'$. In the results section, we show the ROA results obtained by the PS approach for a prototypical molecule, $\bm R$-methyloxirane, and compare them to experiments and the traditional perturbation theory based on the BO states. 

For notation below, three dimensional vectors and operators are bolded. For the most part of the manuscript (except for Eqs.\ref{eq:a2}-\ref{eq:d2} and Appendix \ref{sec_delta_G}), we use the Greek letters $\alpha$, $\beta$, $\gamma$, $\delta$, $\eta$, $\sigma$ in subscripts or superscripts to denote the Cartesian components of a three-dimensional space. We label nuclei with $A, B, C, D$ and quantum states (mostly electronic states) as $I, J, K, L$. For a nuclear vibrational normal mode indexed by $k$, we will label the displacements in position as $\bm X_{k}$; we will label the displacements in momentum space as $\bm Z_{k}$.  Finally, when working with magnetic field problems, there is always the tedious question of the gauge origin; henceforward, this origin will always be denoted $\bG_0$ and should not be confused with $\bG$ or $\bG'$ above. It is obviously essential that all Raman optical activity calculations should be invariant to $\bG_0$ (at least in principle).

\section{Standard Berry Curvature Theory Based on BO Theory}
\label{sec:BO}
\subsection{Raman Optical Activity Tensors}
Let us begin by briefly reviewing the conventional approach to evaluate the ROA tensors in the static limit. Ignoring all nuclear motion, the Hamiltonian describing the interaction of an uncharged system in a time-dependent electric field $\bm F$ and external magnetic field $\bm B$ is:
\begin{align}
    \hat{H} = \hat{H}_{0} - \hat{\bm \mu} \cdot \bm F - \frac{1}{3} \sum_{\alpha\beta} \hat{\theta}_{\alpha\beta} \nabla_{\beta}F_{\alpha} - \hat{\bm m} \cdot \bm B - ...\label{eq:H}
\end{align}
Here $\hat{H}_{0}$ is the electronic Hamiltonian in the absence of external fields and  the raw electric dipole $\hat{\bm \mu}$, electric magnetic dipole $\hat{\bm m}$, and electric quadrupole operators $\hat{\bm \theta}$ are:
\begin{eqnarray}
\label{eq:mu1}
\hat{\mu}_{\alpha} &=& -e(\hat{r}_{\alpha} - G_{0}^{\alpha})\\   
\label{eq:m1}
\hat{m}_{\alpha} &=& -\frac{e}{2m_{e}}\left( (\hat{\bm r}-\bG_0) \times \hat{\bm p} \right)_{\alpha}\\
\label{eq:theta1}
\hat{\theta}_{\beta\gamma} &=& -e\Big(\frac{3}{2}(\hat{\bm r} -\bG_0)_{\beta} (\hat{\bm r} -\bG_0) _{\gamma} - \frac{1}{2}(\hat{\bm r} -\bG_0)^2\delta_{\beta\gamma}\Big) 
\end{eqnarray}
Here $m_e$ denotes the electron mass and $e$ is the positive elementary charge. Note that the matrices defined above formally depend on 
$\bm G_0$ (the origin of space), a topic that will be revisited in detail below. In this regard, most of our focus will be on the electronic magnetic dipole $\bm m$ and electric quadrupole operator $\bm \theta$ for which both on and off-diagonal matrix elements are $\bm G_0$-dependent; the off-diagonal elements of the dipole operator do not depend on $\bm G_0$.

Now,  the raw electronic matrix elements of the above operators are perturbed by the application of an external field, and the responses of the quantum state $\Psi_I$  as
induced by the external fields can be evaluated by time-dependent perturbation theory
\begin{align}
    \mu^{ind}_{\alpha} & = \sum_{\beta}\alpha_{\alpha\beta}F_{\beta} + \sum_{\beta} G_{\alpha\beta}B_{\beta} + \frac{1}{3} A_{\alpha,\beta\gamma} \nabla_{\beta}F_{\gamma} + ... \label{eq:u}\\
    m^{ind}_{\alpha} & = \sum_{\beta} \chi_{\alpha\beta}B_{\beta} + \sum_{\beta} G^{*}_{\beta\alpha}F_{\beta} +... \label{eq:m}\\
    \theta^{ind}_{\alpha\beta} &=  \sum_{\gamma}A_{\gamma,\alpha\beta} F_{\gamma}+ ...\label{eq:theta_roa}
\end{align}
Eqs. \ref{eq:u}-\ref{eq:theta_roa} introduce three key tensors that are  necessary to calculate an ROA signal: the electric-dipole electric-dipole polarizability $\bm \alpha$, the electric-dipole magnetic-dipole polarizability $\bm G$, and the electric-dipole electric quadrupole polarizability $\bm A$. (Note that the notation in Eq. \ref{eq:u} might appear confusing at first, as $\alpha_{\alpha \beta}$ denotes the $\alpha \beta$ component of the polarizability tensor $\bm \alpha$; however, this notation is difficult to avoid in the literature and should be clear from context; we have made sure to avoid this problem in Eqs. \ref{eq:a2}-\ref{eq:d2} and Appendix \ref{sec_delta_G} below, where there might have been some uncertainty given the sheer number of greek letters involved. ) The corresponding sum-over-states expressions from the time-dependent perturbation theory are:
\begin{align}
    {\alpha}_{\alpha\beta} & = \frac{2}{\hbar} \sum_{J\ne I} \frac{\omega_{JI}}{\omega^2_{JI} - \omega^2} \text{Re}\Big[ \bra{\Psi_I} \hat{\mu}_{\alpha}\ket{\Psi_J} \bra{\Psi_J} \hat{\mu}_{\beta}\ket{\Psi_I}\Big]\label{eq:a_real} \\
    {G}'_{\alpha\beta} &= -\frac{2}{\hbar} \sum_{J\ne I} \frac{\omega}{\omega^2_{JI} - \omega^2} \text{Im}\Big[ \bra{\Psi_I} \hat{\mu}_{\alpha}\ket{\Psi_J} \bra{\Psi_J} \hat{m}_{\beta}\ket{\Psi_I} \Big] \label{eq:G_imag}\\
    {A}_{\alpha\beta\gamma} & = \frac{2}{\hbar} \sum_{J\ne I} \frac{\omega_{JI}}{\omega^2_{JI} - \omega^2} \text{Re}\Big[ \bra{\Psi_I} \hat{\mu}_{\alpha}\ket{\Psi_J} \bra{\Psi_J} \hat{\theta}_{\beta\gamma}\ket{\Psi_I}\Big] \label{eq:A_real} 
\end{align}
Here, $\hat{\bm \mu}$, $\hat{\bm m}$, and $\hat{\bm \theta}$ are given above in Eqs. \ref{eq:mu1} - \ref{eq:theta1} . 
$\omega$ is the frequency of the incoming light and $\omega_{IJ}$ is the transition frequency between electronic states $\Psi_I$ and $\Psi_J$.
$\bm G'$ is the imaginary component of the electric-dipole magnetic-dipole polarizability $\bm G$. Note that the imaginary components of $\bm \alpha$ and  $\bm A$ and the real component of $\bm G $ are not included in the ROA tensors because the eigenstates of $\hat{H}_0$ can be chosen to be real within the time-reversal symmetry.

In the static limit ($\omega\rightarrow 0$), the electric-dipole electric-dipole polarizability $\bm \alpha$ in Eq. \ref{eq:a_real},  and electric-dipole electric quadrupole polarizability $\bm A$ in Eq. \ref{eq:A_real} become 
\begin{align}
    {\alpha}_{\alpha\beta}  &= \frac{\partial \bra{\Phi_I} \hat{\mu}_{\beta}\ket{\Phi_I}}{\partial F_{\alpha}} \label{eq:a_dmudF} \\
    {A}_{\alpha\beta\gamma}  &= \frac{\partial \bra{\Phi_I} \hat{\theta}_{\beta\gamma}\ket{\Phi_I}}{\partial F_{\alpha}}\label{eq:A_dthetadF} 
\end{align}
where $\Phi_I$ is the perturbed wavefunction in an external electric field and magnetic field,
\begin{align}
        \ket{\Phi_I} = \ket{\Psi_I} - \sum_{J\ne I}\frac{ \bra{\Psi_J}{   \hat{\bm \mu}\cdot \bm F}\ket{\Psi_I}}{E_I -E_J} \ket{\Psi_J} - \sum_{J\ne I}\frac{ \bra{\Psi_J}{   \hat{\bm m}\cdot \bm B}\ket{\Psi_I}}{E_I -E_J} \ket{\Psi_J} \label{eq:phi}
\end{align}
Eq. \ref{eq:phi} follows for any  hermitian operator from basic perturbation theory, so that $\frac{\partial \bra{\Phi_I} \hat{\mathcal{\bm O}}\ket{\Phi_I}}{\partial F_{\alpha}}$ has a natural sum-over-states form:
\begin{align}
    \frac{\partial \bra{\Phi_I} \hat{\mathcal{\bm O}}\ket{\Phi_I}}{\partial F_{\alpha}} &=  -\sum_{J\ne I}\frac{ \bra{\Psi_I}\hat{ \mu}_{\alpha}\ket{\Psi_J}}{E_I -E_J}\bra{\Psi_J} \hat{\mathcal{\bm O}} \ket{\Psi_I}  - \sum_{J\ne I}\bra{\Psi_I} \hat{\mathcal{\bm O}} \ket{\Psi_J}\frac{ \bra{\Psi_J}\hat{ \mu}_{\alpha}\ket{\Psi_I}}{E_I -E_J} \\
    & = \frac{2}{\hbar} \sum_{J\ne I}\frac{1}{\omega_{JI}}\text{Re}\Big[ \bra{\Psi_I} \hat{\mu}_{\alpha}\ket{\Psi_J} \bra{\Psi_J} \hat{\mathcal{\bm O}}\ket{\Psi_I} \Big]
\end{align}
For the electric-dipole magnetic-dipole polarizability $\bm G'$, one can find an expression without summing over states as follows. As suggested by Amos,\cite{amos_electric_1982} one can divide both sides of Eq. \ref{eq:G_imag} by $\omega$ and then take the static limit. In so doing, one arrives at the Berry-curvature-like expression
\begin{align}
    \omega^{-1}{G}'_{\alpha\beta} &= -\frac{2}{\hbar} \sum_{J\ne I} \frac{1}{\omega^2_{JI}} \text{Im}\Big[ \bra{\Psi_I} \hat{\mu}_{\alpha}\ket{\Psi_J} \bra{\Psi_J} \hat{m}_{\beta}\ket{\Psi_I} \Big]\label{eq:G_sum} \\
    & = -2\hbar\text{Im} \bra{\frac{\partial\Phi_I}{\partial F_{\alpha}}}\ket{\frac{\partial\Phi_I}{\partial B_{\beta}}}\label{eq:G_amos}
\end{align}
To obtain Eq.\ref{eq:G_amos} from Eq.\ref{eq:G_sum}, one can write out the derivatives of the perturbed wavefunction in Eq. \ref{eq:phi} with respect to the external magnetic and electric   field,
\begin{align}
        \ket{\frac{\partial\Phi_I}{\partial  B_{\alpha}}} &= -  \sum_{J\ne I}\frac{ \hat{ m}^{\alpha}_{JI}}{E_I -E_J} \ket{\Psi_J}\label{eq:dphidB} \\
       \bra{\frac{\partial\Phi_I}{\partial  F_{\alpha}}} &= -  \sum_{J\ne I}\frac{ \hat{ \mu}^{\alpha}_{IJ}}{E_I -E_J} \bra{\Psi_J} \label{eq:dphidF}
\end{align}

As mentioned above, in order to calculate non-resonant vibrational Raman optical activity, one must evaluate all three response tensors above when sandwiched between the ground vibrational state and an excited vibrational state.   To that end, it is standard today to invoke the BO approximation, where the total molecular wavefunction is written as a product of nuclear vibrational state and electronic state $\Psi_I(\bm R, \bm r) = \psi_I(\bm r; \bm R) \chi(\bm R)$. Thereafter, for a given vibrational mode $k$, the polarizability tensors can be expanded in the molecular normal mode coordinates $\bm X$ based on  Placzek's approximation. For example, the the electric-dipole magnetic-dipole polarizability $\bm G'$ along normal mode $k$ is defined to first order as: 
\begin{align}
    \omega^{-1}G'^{(k)}_{\alpha\beta}  & \equiv \omega^{-1}\frac{\partial{ G'}_{\alpha\beta}}{\partial  X_{k}}\Big|_{eq} \bra{\chi_0^{(k)}}   X_{k}\ket{\chi_1^{(k)}} \\
        &= \omega^{-1} \sum_{A \gamma} \frac{\partial{ G'}_{\alpha\beta}}{\partial   R_{A \gamma}}\Big|_{ eq} S^{(k)}_{A\gamma}\bra{\chi_0^{(k)}}   X_{k}\ket{\chi_1^{(k)}} \\
    & = \omega^{-1} \sum_{A \gamma} \frac{\partial{ G'}_{\alpha\beta}}{\partial   R_{A \gamma}} \Big|_{ eq} S^{(k)}_{A\gamma} \Big(\frac{\hbar}{2\omega_{k}}\Big)^{\frac{1}{2}} 
    \label{eq:joe}\\
    & = -2\hbar\Big(\frac{\hbar}{2\omega_{k}}\Big)^{\frac{1}{2}}\sum_{A\gamma}\frac{\partial }{\partial  R_{A \gamma}}\text{Im} \bra{\frac{\partial\Phi_I}{\partial F_{\alpha}}}\ket{\frac{\partial\Phi_I}{\partial B_{\beta}}} \Big|_{eq} S^{(k)}_{A\gamma}\label{eq:dGdq} 
\end{align}
Here, $\bm S^{(k)} = \frac{\partial \bm R}{\partial X_k} \Big |_{eq}$ describes the Cartesian nuclear displacements in normal mode $k$ and   
\begin{align}
    \bra{\chi_0^{(k)}}X_{k}\ket{\chi_1^{(k)}} = \bra{\chi_1^{(k)}} X_{k}\ket{\chi_0^{(k)}} = \Big(\frac{\hbar}{2\omega_{k}}\Big)^{\frac{1}{2}}
\end{align}
is the standard mass-weighted position operator matrix elements for a harmonic oscillator. 

\subsection{Circular Intensity Difference}
Experimental ROA intensities routinely measure the dimensionless circular intensity difference (CID)
\begin{align}
    \Delta_{\alpha} = \frac{I^{R}_{\alpha}-I^{L}_{\alpha}}{I^{R}_{\alpha}+I^{L}_{\alpha}},
\end{align}
where $I^{R}_{\alpha}$ and $I^{L}_{\alpha}$ are the scattered Raman intensities for a molecule subjected  to right- and left-circularly polarized excitation, respectively, and the scattered light is analyzed in a linear polarization defined by the analyzer angle $\alpha$ with respect to the chosen reference. \cite{barron_experimental_1991}. The  CID expressions for different experimental setups with incident light of wavelength $\lambda$ have been listed in terms of ROA invariants in Ref. \citenum{polavarapu_ab_1990} and are also listed in Table \ref{tab:CID} here.

\renewcommand{\arraystretch}{1.5}
\begin{table}[htbp]
\centering 
\caption{Relevant expressions for circular intensity difference  under different measurement types.} \label{tab:CID}
\begin{tabular} {c c}
\textbf{Measurement type} & $\bm \Delta = \frac{\bm I^{R}-\bm I^{L}}{\bm I^{R}+\bm I^{L}}$ \\
\hline
Depolarized &
$\displaystyle \frac{4\pi}{\lambda} \left[ \frac{\gamma^2}{\omega} - \frac{1}{3} \frac{\delta^2}{\omega} \right]/\beta^2$ \\
\hline
Polarized &
$\displaystyle \frac{4\pi}{\lambda} \left[ \frac{45}{\omega} \alpha G'+ \frac{7}{\omega} \gamma^2 + \frac{1}{\omega} \delta^2 \right]/\left[45\alpha^2+7\beta^2\right]$ \\
\hline
Magic angle &
$\displaystyle \frac{4\pi}{\lambda} \left[ \frac{9}{\omega} {\alpha}{G}' + \frac{2}{\omega} \gamma^2 \right]/\left[9\alpha^2+2\beta^2\right]$ \\
\hline
Backward &
$\displaystyle \frac{48\pi}{\lambda}\left[ \frac{1}{\omega} \gamma^2 + \frac{1}{3\omega} \delta^2 \right]/\left[45\alpha^2+7\beta^2\right]$ \\
\hline
Forward &
$\displaystyle \frac{48\pi}{\lambda}\left[ \frac{45}{\omega} {\alpha} {G}' + \frac{1}{\omega} \gamma^2 - \frac{1}{\omega} \delta^2 \right]/\left[45\alpha^2+7\beta^2\right]$ \\
\hline
\end{tabular}
\end{table}

In Table \ref{tab:CID} above, several
 ROA invariants have been calculated from the polarizabilities in Eqs. \ref{eq:a_dmudF}, \ref{eq:A_dthetadF}, and \ref{eq:G_amos}.
\begin{align}
    \alpha^2 &= \frac{1}{9} \sum_{ji} \alpha_{jj}\alpha_{ii}\label{eq:a2} \\
    \beta^2 & = \frac{1}{2} \sum_{ji} (3\alpha_{ji}\alpha_{ji} - \alpha_{jj}\alpha_{ii})\\
    \alpha G' &= \frac{1}{9} \sum_{ji} \alpha_{jj} G'_{ii}\label{eq:ag}\\
    \gamma^2 & = \frac{1}{2}\sum_{ji} (3\alpha_{ji}G'_{ji} - \alpha_{jj}G'_{ii})\label{eq:g2} \\
    \delta^2 & = \frac{1}{2}\omega \sum_{ji} \alpha_{ji}\sum_{ln}\epsilon_{jln}A_{lni}\label{eq:d2} 
\end{align}
Note that above, in Eqs. \ref{eq:a2}-\ref{eq:d2}, we have used $i,j,l,n$ to index three-dimensional cartesian unit vectors (in order to avoid double counting the greek variables on the left hand side); in general, however,   we will continue to use the subscripts or superscripts $\alpha, \beta, \gamma, \eta, \sigma$ to index the three cartesian directions.
In order to calculate ROA signals using Table \ref{tab:CID}, all of the invariants in Eqs. \ref{eq:a2}-\ref{eq:d2} must be expanded in terms of normal modes, as in Eq. \ref{eq:dGdq} above for $\omega^{-1}\bm G'^{(k)}$.

A further is comment is now appropriate regarding the choice of gauge origin. Although the polarizability tensor $\alpha_{ij}$ is invariant to the choice of origin $\bm G_0$, the same is not true for either the electric-dipole magnetic-dipole polarizability $\bm G'$ or the electric-dipole electronic-quadrupole polarizability tensor $\bm A$; these tensors do depend on the choice of origin (through the factors of $\hat{\bm m}$ and $\hat{\bm \theta}$ in Eqs. \ref{eq:G_imag} and  \ref{eq:A_real} above).  That being said, however, the expressions $\alpha G'$ and $\gamma^2$ in Eqs. \ref{eq:ag} and \ref{eq:g2} {\em are} invariant to the gauge origins within BO theory. This invariance can be ascertained by noting that within BO theory, 
\begin{align}
\label{eqn:rp_same}
    \left[ \hH_{BO},\hat{\br} \right] = -i\hbar \frac{\hat{\bp}}{m_e},
\end{align}
which implies $\bra{\Psi_{I}} \hat{\bm \mu}\ket{\Psi_J} \propto -e\bra{\Psi_{I}}  \hat{\bm p}\ket{\Psi_J} $ for electronic states $I$ and $J$. Invariance of $\omega^{-1} \sum_{i} G'_{ii}$ in the static limit follows by noting that the trace is the dot product of ${\bm \mu}_{IJ}$ and ${\bm m}_{JI}$ in  Eq. \ref{eq:G_imag}, which vanishes because $\hat{\bm m}$ is  a cross product of $\hat{\bm r}$ with $\hat{\bm p}$ as in Eq. \ref{eq:m1}; furthermore, the same vanishing also holds for the dot product of $ \partial {\bm \mu}_{IJ}/{\partial \bR}$ and $ \partial {\bm m}_{JI}/{\partial \bR}$. For a discussion of the invariance of $\delta^2$, see Appendix \ref{sec_delta_G}.

Finally, for magic-angle measurements, the predicted ROA CID (according to Table \ref{tab:CID})  
is:
\begin{align}
    \Delta^{(k)}_{BO}(*) 
    & = \frac{4\pi}{\lambda} \left[\sum_{\alpha\beta }  \frac{\partial \alpha_{\alpha\beta}}{\partial X_{k}} \frac{\omega_k^{-1} \partial G'_{\alpha\beta}}{\partial X_{k}} \right]
    \Big/\left[\sum_{\alpha\beta }\frac{\partial \alpha_{\alpha\beta}}{\partial X_{k}} \frac{\partial \alpha_{\alpha\beta}}{\partial X_{k}}\right]\label{eq:CIDm_BO}
\end{align}
Here, if we work with vibration $k$, we have replaced $\alpha_{\alpha \beta}$ with $\frac{\partial \alpha_{\alpha \beta}}{\partial X_k} \bra{\chi_0^{(k)}}   X_{k}\ket{\chi_1^{(k)}}$, etc.
Similar formulas can be found for the other CID expressions. Henceforward, given that 
Table \ref{tab:CID} involves  
$\omega^{-1}\bm G'$, for which there is a the Berry curvature expression (Eq. \ref{eq:G_amos}), we will refer to Table \ref{tab:CID} above as either  BO or Berry curvature ROA expressions.

\section{Phase Space Theory of ROA}
\label{sec:PS}

As far as predicting ROA signals, the key element above is the $\omega^{-1}\frac{ \partial G'_{\gamma\alpha}}{\partial  X}$ term in Eq. \ref{eq:CIDm_BO}. Interestingly, 
the matrix elements of $\omega^{-1}\partial G'_{\gamma\alpha}/\partial \bm R$ can be reintrepreted through phase space theory, which we will now discuss.

\subsection{Brief Review of Shenvi's Phase Space Electronic Structure Theory}
\label{sec:shenvi}
As was established long ago by many chemists (including Nafie\cite{nafie_adiabatic_1983}, Stephens\cite{Stephens_1985_MFP},  and Patchkovski\cite{Patchkovskii2012}), in the context of chiroptical spectroscopy, the main failure of BO theory is the lack of any dependence of the nuclear momentum on the electronic wave function. After all, within the BO approximation, one discards the term
\begin{eqnarray}
    \hat{H}' =   - i\hbar\sum_{IJ}\frac{\bP\cdot  {{\bm d}}_{IJ} }{M}\ket{\psi_I}\bra{\psi_J} 
\label{eq:nonBO}
\end{eqnarray}
Here $ {{\bm d}}_{IJ} = \bra{\psi_{I}}\frac{\partial}{\partial \bm R} \ket{\psi_J}$ is the derivative coupling vector and $\hat{H}'$ can be considered the first-order correction to BO theory.
As a result of ignoring the coupling in Eq. \ref{eq:nonBO} above BO lacks electronic momentum\cite{Takatsuka_timeshiftflux_2009,Bian2023,Polkovnikov2025} and cannot predict VCD theory directly (among other ailments).\cite{duston_phase-space_2024,nafie_adiabatic_1983,Patchkovskii2012,tao_practical_2024}  

Now, in 2008, Shenvi  tried to incorporate this nuclear momentum dependence back to electronic states by diagonalizing the molecular Hamiltonian in the adiabatic basis to obtain superadiabatic states.\cite{shenvi_phase-space_2009}
   \begin{align} 
    \hat{H}_{\rm Shenvi}(\bm{R},\bP) 
    = &\sum_{IJK,A} \frac{1}{2M_A} \left( \bm{P}_A \delta_{IJ} - i\hbar {{\bm d}}^A_{IJ} \right)\cdot
    \left( \bP_A \delta_{JK} - i \hbar {{\bm d}}^A_{JK} \right)\ket{\psi_I}\bra{\psi_K} 
     +\sum_K E_{K}(\bm{R})\ket{\psi_K}\bra{\psi_K}, 
    \label{eq:shenvi} 
\end{align}
 This formalism, however, is not practically stable due to the well-known diverging character of the derivative coupling vectors near avoided crossings and high computational expense of computing derivative couplings in a select subspace of adiabatic states. Furthermore, derivative couplings are known to be gauge dependent, which are not well defined in the presence of degenerate states.
Despite these failures, however, it is worth noting that Shenvi's approach does provide an interpretation of Eq. \ref{eq:G_amos} above, which we now discuss.

\subsection{Interpretation of $ G'$ tensor  
in terms of Nuclear Momentum Dependence}
Let us begin by recapitulating\cite{nafie_adiabatic_1983,duston_phase-space_2024} how and why an electronic wavefunction depends on nuclear momentum, and how such dependence allows us to calculate the AAT (Eq. \ref{eq:AAT}) for VCD calculations; thereafter, we will show how one can similarly express $\omega^{-1}\frac{ \partial G'_{\gamma\alpha}}{\partial \bm R}$ in terms of derivatives with respect to nuclear momentum within phase space theory. 
If one considers the term $ - i\hbar\sum_{IJ}\frac{\bP\cdot  {{\bm d}}_{IJ} }{M}\ket{\psi_I}\bra{\psi_J} $ as a perturbation on top of BO theory, then within phase space electronic structure theory,  the nuclear momentum dependence of the electronic wavefunction can also be written in a perturbative way:
\begin{align}
\label{eq:psi}
    \ket{\Phi^{\text{P}}_I} &= \ket{\psi_I} - i\hbar \sum_{J\ne I}\frac{\sum_A \frac{\bP^A}{M_A}  \cdot {\bm d}^A_{JI}}{E_I -E_J} \ket{\psi_J} \\
    & = \ket{\psi_I} - i\hbar \sum_{J\ne I}\frac{\sum_A \frac{\bP^A}{M_A}  \cdot\bra{\psi_J} \frac{\partial \hat{H}}{\partial \bm R_A}\ket{\psi_I}}{(E_I -E_J)^2} \ket{\psi_J} \label{eq:phi_P}
\end{align}
Thus, if we differentiate an electronic wavefunction derivative with respect to the nuclear momentum, we find:
\begin{align}
    \ket{\frac{\partial\Phi^{\text{P}}_I}{\partial \bm P_{A}}} &= - i\hbar \sum_{J\ne I}  \frac{1}{M_A} \frac{  {\bm d}^A_{JI}}{E_I -E_J} \ket{\psi_J} = - i\hbar \sum_{J\ne I}  \frac{1}{M_A} \frac{ \bra{\psi_J} \frac{\partial \hat{H}}{\partial \bm R_A}\ket{\psi_I}}{(E_I -E_J)^2} \ket{\psi_J}\label{eq:dphidP}
\end{align}
Similarly, we can write down the dependence of the electronic wavefunction on the nuclear displacement
\begin{align}
        \ket{\Phi_I^{\text{R}}} &= \ket{\psi_I} + \sum_{J\ne I}\frac{\sum_{A\alpha}\bra{\psi_J} \frac{\partial \hat{H}}{\partial R_{A\alpha}}\ket{\psi_I}\delta R_{A\alpha}}{E_I -E_J} \ket{\psi_J}  
\end{align}        
and the wavefunction derivative with respect to the nuclear displacement
\begin{align}
        \ket{\frac{\partial\Phi^{\text{R}}_I}{\partial  R_{A\alpha}}} & = \sum_{J\ne I}\frac{\bra{\psi_J} \frac{\partial \hat{H}}{\partial R_{A\alpha}}\ket{\psi_I}}{E_I -E_J} \ket{\psi_J}  \label{eq:dphidR}
\end{align}
Interestingly,  from Eqs.  \ref{eqn:rp_same}, \ref{eq:dphidP} and \ref{eq:dphidR},\cite{duston_phase-space_2024} one can establish Nafie's relationship\cite{nafie_adiabatic_1983},
\begin{eqnarray}
    \label{eqn:nafie}
    \langle \hat{\bm p} \rangle = m_e \frac{d     \langle \hat{\bm r} \rangle}{dt}
 \; \; \; \mbox{or equivalently} \; \; \; 
    -e\frac{\partial \bm \langle \hat{\bm p}\rangle}{\partial P_{B\beta}} \Big|_{\bm P =0}  = \frac{m_e}{M_B} \frac{\partial \langle\hat{\bm \mu}\rangle}{\partial  R_{B\beta}}\Big|_{\bm R =0}  
    \label{eq:APT_equiv}
\end{eqnarray}
which will be important below in Sec. \ref{sec:PS_W}.

In any event, for our present purposes, combining Eq.\ref{eq:dphidB}, Eq. \ref{eq:dphidP}, and Eq.\ref{eq:dphidR}, we can derive the AAT expression based on the magnetic field perturbation theory\cite{Stephens_1985_MFP}
\begin{align}
    \frac{\partial \bra{\Phi^{\text{P}}_I}\hat{ m}_{\alpha}\ket{\Phi^{\text{P}}_I}}{\partial  P_{A\beta}} &= \left<\frac{\partial\Phi^{\text{P}}_I}{\partial  P_{A\beta}}\middle|\hat{ m}_{\alpha}\middle|\Phi^{\text{P}}_I\right> +  \left<\Phi^{\text{P}}_I \middle| \hat{ m}_{\alpha}\middle|\frac{\partial\Phi^{\text{P}}_I}{\partial  P_{A\beta}}\right> \\
    & = -i\hbar \sum_{J\ne I}  \frac{1}{M_A} \frac{ \bra{\psi_I} \frac{\partial \hat{H}}{\partial  R_{A\beta}}\ket{\psi_J}}{(E_I -E_J)^2} \bra{\psi_J}\hat{ m}_{\alpha}\ket{\psi_I} - i\hbar \sum_{J\ne I}  \frac{1}{M_A} \frac{ \bra{\psi_J} \frac{\partial \hat{H}}{\partial  R_{A\beta}}\ket{\psi_I}}{(E_I -E_J)^2} \bra{\psi_I}\hat{ m}_{\alpha}\ket{\psi_J} \\
    & = \frac{2\hbar}{M_A}\text{Im}\Bigg[ \sum_{J\ne I} \frac{ \bra{\psi_J} \frac{\partial \hat{H}}{\partial R_{A\beta}}\ket{\psi_I}}{(E_I -E_J)^2} \bra{\psi_I}\hat{ m}_{\alpha}\ket{\psi_J}\Bigg] \\
    &=-\frac{2\hbar}{M_A}\text{Im} \bra{\frac{\partial\Phi_I}{\partial B_{\alpha}}}\ket{\frac{\partial\Phi_I}{\partial R_{A\beta}}} \label{eq:AAT}
\end{align}
Differentiating the AAT with respect to the external electric field yields
\begin{align}
    \frac{\partial^2 \bra{\Phi^{\text{P}}_I}\hat{ m}_{\alpha}\ket{\Phi^{\text{P}}_I}}{\partial F_{\gamma}\partial P_{A\beta}}  = -\frac{2\hbar}{M_A}\frac{\bm \partial}{\partial F_{\gamma}}\text{Im} \bra{\frac{\partial\Phi_I}{\partial B_{\alpha}}}\ket{\frac{\partial\Phi_I}{\partial R_{A\beta}}} \label{eq:dmdFdP}
\end{align}
Note that one must go beyond the BO approximation to include electron–nuclear momentum coupling in Eq.\ref{eq:phi_P} to find a non-zero AAT.\cite{nafie_adiabatic_1983,duston_phase-space_2024}

Similarly, from the same wavefunction perturbations in Eq.\ref{eq:dphidF}, Eq. \ref{eq:dphidP}, and Eq.\ref{eq:dphidR}, one can write down 
\begin{align}
     \frac{\partial^2 \bra{\Phi^{\text{P}}_I}\hat{ \mu}_{\gamma}\ket{\Phi^{\text{P}}_I}}{\partial B_{\alpha}\partial  P_{A\beta}}  =  -\frac{2\hbar}{M_A}\frac{\bm \partial}{\partial B_{\alpha}}\text{Im} \bra{\frac{\partial\Phi_I}{\partial F_{\gamma}}}\ket{\frac{\partial\Phi_I}{\partial R_{A\beta}}}
    \label{eq:dmudBdP}
\end{align}
As above, this term will be zero within the BO approximation as there is no nuclear momentum dependence of the electronic wavefunction even in the presence of an external magnetic field.

Now we introduce a new definition. 
\begin{align}
    W_{\gamma\alpha}(\bR,\bP) = \frac{\partial \bra{\Phi^{\text{P}}_I}\hat{ \mu}_{\gamma}\ket{\Phi^{\text{P}}_I}}{\partial B_{\alpha}} -\frac{\partial \bra{\Phi^{\text{P}}_I}\hat{ m}_{\alpha}\ket{\Phi^{\text{P}}_I}}{\partial F_{\gamma}} 
\label{eq:W}
\end{align}
Note that the wavefunction $\Phi^{\text{P}}_I$ is the perturbed wavefunction arising from the perturbing nonadiabatic operator $-i\hbar{\frac{\bm P}{M}}\cdot{\hat{\bm d}}$, which reduces to BO wavefunction when $\bm P=0$. The newly introduced tensor $W_{\gamma\alpha}$ is only non-zero at $\bm P\ne 0$ and represents the asymmetry of the second-order electronic responses when the nuclei are moving (so as to break time-reversal symmetry). In other words, within BO theory (which is nearly equivalent to phase space theory when $\bm P=0$ up to a second-order term in $\bm d_{IJ}$), for arbitrary $\bm B$ and $\bm F$ fields and position $\bm R$, 
\begin{align}
   \bra{\psi}\hat{\bm m} \ket{\psi} &=-\bra{\psi}\frac{\partial\hat{H}_{BO}}{\partial \bm B} \ket{\psi} = - \frac{\partial\bra{\psi}\hat{H}_{BO}\ket{\psi}}{\partial \bm B}\approx -\frac{\partial \left< \hat{H}_{Shenvi}\right>}{\partial \bm B} \Big |_{\bm P=0}\label{eq:dH_dB} \\
    \bra{\psi}\hat{\bm \mu} \ket{\psi} &= -\bra{\psi}\frac{\partial\hat{H}_{BO}}{\partial \bm F} \ket{\psi} =-\frac{\partial\bra{\psi}\hat{H}_{BO}\ket{\psi}}{\partial \bm F} \approx -\frac{\partial \left< \hat{H}_{Shenvi} \right>}{\partial \bm F}  \Big |_{\bm P=0}\label{eq:dH_dF}
\end{align}
so that the $\bm W$ tensor must be exactly zero:
\begin{align}
    \frac{\partial \bra{\psi}\hat{ \mu}_{\gamma}\ket{\psi}}{\partial B_{\alpha}} = - \frac{\partial^2\bra{\psi}\hat{H}_{BO}\ket{\psi}}{\partial F_{\gamma}\partial B_{\alpha}}=\frac{\partial \bra{\psi}\hat{ m}_{\alpha}\ket{\psi}}{\partial F_{\gamma}}\label{eq:W_BO} 
\end{align}
However, while we are free to differentiate Eq. \ref{eq:W_BO} with respect to nuclear positions $\bm R$, so that 
\begin{align}
    \frac{\partial^2 \bra{\psi}\hat{ \mu}_{\gamma}\ket{\psi}}{\partial R_{A\beta}\partial B_{\alpha}} =\frac{\partial^2 \bra{\psi}\hat{ m}_{\alpha}\ket{\psi}}{\partial R_{A\beta}\partial F_{\gamma}} 
\end{align}
we cannot take such derivatives with respect to nuclear momentum $\bm P$ (where we necessarily must go beyond BO theory):
\begin{align}
    \frac{\partial^2 \bra{\Phi^{P}}\hat{ \mu}_{\gamma}\ket{\Phi^{P}}}{\partial P_{A\beta}\partial B_{\alpha}} \ne\frac{\partial^2 \bra{\Phi^{P}}\hat{ m}_{\alpha}\ket{\Phi^{P}}}{\partial P_{A\beta}\partial F_{\gamma}} 
\end{align} In other words, if $\bP \ne 0$, 
\begin{align}
    \frac{\partial\bra{\Phi^{P}}\hat{ \mu}_{\gamma}\ket{\Phi^{P}}}{\partial B_{\alpha}} \ne\frac{\partial \bra{\Phi^{P}}\hat{ m}_{\alpha}\ket{\Phi^{P}}}{\partial F_{\gamma}} 
\end{align}
These relationships break down  because, by allowing electrons to move with nuclei, a nonzero electronic angular momentum ($\bm m$) can arise both from internal nuclear motion $\bm P$ and an external magnetic field $\bm B$, so that Eq. \ref{eq:W_BO} holds only when $\bm P = 0$.
See below, Sec. \ref{sec:PS_W}, for a complete analysis of this term within phase space theory. 

Taking the derivative of $ W_{\gamma\alpha}$ with respect to nuclear momentum, one arrives at the  the ROA tensor $\omega^{-1}\frac{ \partial G'_{\gamma\alpha}}{\partial \bm R}$
\begin{align}
    \frac{\partial  W_{\gamma\alpha}}{\partial \bm P_{\beta}}=&\frac{\partial^2 \bra{\Phi^{\text{P}}_I}\hat{ \mu}_{\gamma}\ket{\Phi^{\text{P}}_I}}{\partial B_{\alpha}\partial \bm P_{\beta}} -\frac{\partial^2 \bra{\Phi^{\text{P}}_I}\hat{ m}_{\alpha}\ket{\Phi^{\text{P}}_I}}{\partial F_{\gamma}\partial \bm P_{\beta}} \label{eq:dGdp_diff}\\
   =  & \frac{2\hbar}{\bm M}\text{Im}\left[\bra{\frac{\partial\Phi_I}{\partial B_{\alpha}}}\ket{\frac{\partial^2\Phi_I}{\partial F_{\gamma}\partial \bm R_{\beta}}}-\bra{\frac{\partial\Phi_I}{\partial F_{\gamma}}}\ket{\frac{\partial^2\Phi_I}{\partial B_{\alpha}\partial\bm  R_{\beta}}}\right] \\
    = & \frac{2\hbar}{\bm M}\text{Im}\left[\bra{\frac{\partial\Phi_I}{\partial B_{\alpha}}}\ket{\frac{\partial^2\Phi_I}{\partial F_{\gamma}\partial \bm R_{\beta}}}+\bra{\frac{\partial^2\Phi_I}{\partial B_{\alpha}\partial \bm R_{\beta}}}\ket{\frac{\partial\Phi_I}{\partial F_{\gamma}}}\right]\\
=&  \frac{2\hbar}{\bm M}\frac{\partial}{\partial \bm R_{\beta}}\text{Im} \bra{\frac{\partial \Phi_I}{\partial B_{\alpha}}}\ket{\frac{\partial\Phi_I}{\partial F_{\gamma}}} \\
    = & \frac{1}{\bm M} \omega^{-1}\frac{ \partial G'_{\gamma\alpha}}{\partial \bm R_{\beta}} \label{eq:ps_ROA_tensor}
\end{align}

Using Eqs. \ref{eq:joe} and \ref{eq:ps_ROA_tensor}, it now follows that we can express $G'$ along mode $k$ directly in terms of momentum derivatives:
\begin{align}
\omega^{-1}G'^{(k)}_{\alpha\beta} 
  & = \omega^{-1} \sum_{A \gamma} \frac{\partial{ G'}_{\alpha\beta}}{\partial   R_{A \gamma}} \Big|_{ eq} S^{(k)}_{A\gamma} \Big(\frac{\hbar}{2\omega_{k}}\Big)^{\frac{1}{2}} \\
    & = \sum_{A \gamma} M_A\frac{\partial  W_{\alpha\beta}}{\partial  P_{A\gamma}} S^{(k)}_{A\gamma} \Big(\frac{\hbar}{2\omega_{k}}\Big)^{\frac{1}{2}}\label{eq:new0}  
    \\
    & \equiv \frac{\partial W_{\alpha\beta}}{\partial Z_k}\Big(\frac{\hbar}{2\omega_{k}}\Big)^{\frac{1}{2}}  
    \label{eq:new1}
\end{align}
The meaning of Eq. \ref{eq:new1}
is as follows.
For a normal model with displacements $\bm S^{(k)}$ and mass-weighted phase space coordinates $(X_k,Z_k)$,
the electric-dipole magnetic-dipole polarizability $\bm G'$ along mode $k$ (as defined with normal mode position coordinates $X_k$ in Eq.\ref{eq:dGdq}) can be calculated alternatively by evaluating the changes with respect to the momentum coordinates $ Z_k$.
Intuitively, a factor of $M_A$ arises because $\frac{\partial P_{A\alpha}}{\partial Z_k} = M_A S^{(k)}_{A\alpha}$.

Hence, along normal mode $k$, the $\bm G'$ tensor, and the CID signals can be calculated with both approaches (Eq. \ref{eq:dGdq} and Eq. \ref{eq:new1}). As an example, for magic angle CID, for a given vibrational transition from ground to kth vibrational state, the result is (compare with Eq. \ref{eq:CIDm_BO}):

\begin{align}
       \Delta^{(k)}_{PS}(*) 
    & = \frac{4\pi}{\lambda} \left[\sum_{\alpha\beta }  \frac{\partial \alpha_{\alpha\beta}}{\partial X_{k}}  \frac{\partial   W_{\alpha \beta} }{\partial{ Z}_{k}} \right]
    \Big/\left[\sum_{\alpha\beta }\frac{\partial \alpha_{\alpha\beta}}{\partial X_{k}} \frac{\partial \alpha_{\alpha\beta}}{\partial X_{k}}\right]\label{eq:CIDm_PS}
\end{align}

\subsection{A More Practical Theory of Phase Space Electronic Structure Theory}

The theory above makes clear that ROA can be explained naturally in the context of an electronic structure theory paramaterized by both $\bm R$ and $\bm P$.
That being said, at the end of Sec. \ref{sec:shenvi}, we argued that the Shenvi approach in Eq. \ref{eq:shenvi} was not optimal.
Thus, to make further progess, one must be prepared to make approximations. 
To that end, over the last few years, 
our research groups have proposed an electronic phase-space Hamiltonian, where we approximate the non-diverging part of the derivative coupling with a one-electron $\hat{\bm \Gamma}$ operator. \cite{tao_basis-free_2025}
\begin{align}
\hH_{\rm PS}(\bm{R},\bP) &= \sum_{A}\frac{1}{2M_A} \left( \bm{P}_A  - i\hbar \hbm{\Gamma}_A(\bR) \right)\cdot
    \left( \bP_A - i \hbar \hbm{\Gamma}_A(\bR) \right)+  \hH_{el}(\bm{R}),\label{eq:PSH} 
\end{align}
The nuclear-momentum coupling term $-i\hbar\frac{\bm P}{\bm M}\cdot \hat{\bm \Gamma}$ can be viewed as the surrogate first-order correction\cite{wu_recovering_2025} beyond the BO approximation, provided $\hat{\bGamma}$ approximates the derivative coupling operator. Unlike Shenvi's approach, which requires diagonalizing electronic Hamiltonian to obtain adiabatic states, we directly diagonalize the phase space electronic Hamiltontian given in Eq. \ref{eq:PSH}, $\hat{H}_{PS}\ket{\psi_{PS}}=E_{PS}\ket{\psi_{PS}}$. The key question is then how to construct $\hat{\bGamma}$ exactly. Although introducing approximation could admittedly invalidate Eq. \ref{eqn:nafie} above, so that
\begin{eqnarray}
\label{eqn:no_nafie}
    -e\frac{\partial \bra{\psi_{PS}}\hat{\bm p}\ket{\psi_{PS}}}{\partial P_{B\beta}} \Big|_{\bm P =0}  \ne \frac{m_e}{M_B} \frac{\partial \bra{\psi_{PS}}\hat{\bm \mu}\ket{\psi_{PS}}}{\partial  R_{B\beta}}\Big|_{\bm R =0}  
\end{eqnarray}
the resulting hamiltonian with an approximate form of $\hat{\bGamma}$ can be extremely useful (and stable).  That being said, the inapplicability of Eq. \ref{eqn:nafie} will have consequences that we will need to address below.

\subsubsection{Field Free Phase Space Electronic Structure Theory}

In the absence of a magnetic field,  the derivative coupling operator  $ {{ d}}^A_{JK}$  satisfies (for $J \ne K$) the symmetries listed below in Eqs. \ref{eq:dconstrain1a}-\ref{eq:dconstrain4a}.\cite{Wu2022} 
\begin{eqnarray}
    -i\hbar\sum_{A} { d}^A_{JK} + \bra{\psi_J}\bm{\hat{p}} \ket{\psi_K} &=& 0,\label{eq:dconstrain1a}  \\
    \sum_I \bm \nabla_{A} { d}^A_{JK}  &=& 0, 
    \label{eq:dconstrain2a}\\
    -i\hbar\sum_{A}{\bm R}_{A} \times {d}^A_{JK} + \bra{\psi_J} \bm{\hat{l}} + \bm{\hat{s}} \ket{\psi_K} &=& 0,\label{eq:dconstrain3a}\\
     -\sum_A \left({\bm R}_A \times \bm \nabla_{A} {d}^{B\beta}_{JK}\right)_{\alpha} &
     =&  \sum_{\gamma} \epsilon_{\alpha \beta \gamma} d^{B\gamma}_{JK},\label{eq:dconstrain4a} 
\end{eqnarray}
Because there is no effect when we move electrons and nuclei together,  Eq. \ref{eq:dconstrain1a} shows that  the result of summing the derivative coupling vector between states $J$ and $K$ over all nuclei is equals to the electronic linear momentum between those two states  ${\bm p}_{JK}$. A similar relation is presented in Eq. \ref{eq:dconstrain3a} for a rigid rotation which equates the total nuclear angular momentum to the total electronic
angular momentum operator $\hat{\bm l}+ \hat{\bm s}$. Of course, the derivative coupling is not well-defined for $J=K$ and really amounts to a choice of phase convention. As argued by Littlejohn {\em et al},\cite{Littlejohn2023} the most natural phase convention is to choose:
\begin{align}
   \Big( \hat{\bm P} + \hat{\bm p} \Big) \psi (\bm r, \bm R)   &=0 \label{eq:BO_Pp}\\
     \Big( \hat{\bm J} + \bm{\hat{l}} + \bm{\hat{s}} \Big) \psi (\bm r, \bm R) \label{eq:BO_Ll}  &=0 
\end{align}
which are automatically consistent with Eqs. \ref{eq:dconstrain1a} and \ref{eq:dconstrain3a} for $J \ne K$. Here $\hat{\bm P}  = -i\hbar \sum_{A}\frac{\partial}{\partial \bm R_A}$ and $\hat{\bm J} = -i\hbar \sum_{A}\bm R_A\times\frac{\partial}{\partial \bm R_A}$.
Altogether, if we now extrapolate these symmetry relations for a one-electron operator $\hat{\bGamma}$, we postulate the resulting operator should satisfy the following corresponding symmetries:\cite{tao_basis-free_2025} 

\begin{align}
    -i\hbar\sum_{A}\hbm{\Gamma}_{A} + \hbm{p} &= 0,\label{eq:Gamma1}  \\
    \Big[-i\hbar\sum_{B}\pp{}{\bm{R}_B} + \hbm{p}, \hbm{\Gamma}_A\Big] &= 0,\label{eq:Gamma2}\\
    -i\hbar\sum_{A}{\bm R}_{A} \times \hat{\bm \Gamma}_{A} + \hbm{l} + \hbm{s} &= 0,\label{eq:Gamma3}\\
     \Big[-i\hbar\sum_{B}\left(\bm{R}_B \times\pp{}{\bm{R}_B}\right)_{\gamma} + \hat{l}_{\gamma} + \hat{s}_{\gamma}, \hat{\Gamma}_{A \delta}\Big] &
     = i\hbar \sum_{\alpha} \epsilon_{\alpha \gamma \delta} \hat{\Gamma}_{A \alpha},\label{eq:Gamma4}
\end{align} 
The constraints in Eqs. \ref{eq:Gamma1} and \ref{eq:Gamma3} partition the electronic linear momentum and angular momentum to each moving nucleus to conserve total (nuclear+electronic) linear and angular momentum. In this paper, we focus on systems lacking net spin and therefore omit the spin operator $\hat{\bm s}$ from the definition of $\hat{\bm \Gamma}$ henceforth.  The constraints in Eqs. \ref{eq:Gamma2} and \ref{eq:Gamma4} ensure the phase space energy is invariant to molecular translation and rotation, respectively.\cite{tao_basis-free_2025} 

Following these symmetry constraints, we have previously designed a one-electron $\hat{\bm \Gamma}$ operator\cite{tao_basis-free_2025} consisting of an electron translational factor (ETF) $\hat{\bm \Gamma'}$\cite{schneiderman:1969:pr:etf,Fatehi2011} and an electron rotational factor (ERF)  $\hat{\bm \Gamma''}$ . \cite{athavale2023,qiu_simple_2024}
\begin{eqnarray}
   {\hat{ \bm\Gamma}}_A &=&{\hat{ \bm\Gamma}}_A' +{\hat{ \bm\Gamma}}_A''\label{eq:gammasum}\\
    {\hat{ \bm\Gamma}}_A' &=& \frac{1}{2i\hbar}\left( \hat{\Theta}_A(\hbr){\hat{\bm{ p}}}+{\hat{\bm{ p}}}\hat{\Theta}_A(\hbr)\right)\label{eq:etf}\\
     \hat{\bm{\Gamma}}_A^{''} &=& \sum_{B}  \zeta_{AB}\left(\bm R_A -\bm R^0_{B}\right)\times \left(\bm{K}_B^{-1}\frac{1}{2i\hbar}(\hbm{r}-\bm R_B)\times(\hat{\Theta}_B(\hbr) \hat{\bm{p}} + \hat{\bm{ p}}\hat{\Theta}_B(\hbr))\right)\label{eq:erf}
\end{eqnarray}
The ETF term involves a partition-of-unity operator $\hat{\Theta}$ in the real three-dimensional space to partition the electron-nuclear coupling regions under the translational-type of nuclear motion. 
\begin{align}
\label{eq:theta}
    \hat{\Theta}_A(\hbm{r}) = \frac{eQ_A\mathrm{e}^{-|\hbm{r}-\bm{R}_A|^2/\sigma^2}}{\sum_B eQ_B\mathrm{e}^{-|\hbm{r}-\bm{R}_B|^2/\sigma^2}}.
\end{align}
Here the prefactors $Q_A$ are the bare nuclear charges of nuclei A and $\sigma$ is a locality parameter that controls the width of the electronic-nuclear momentum couplings. The ERF term ${\hat{\bm \Gamma}}''$ includes another locality function $\bm \zeta$ that partitions the regions of coupling for the rotational-type of motion with a locality parameter $\beta$. 
\begin{align}
\label{eq:zeta}
   \zeta_{AB} =  M_{A}e^{-|\bm{R}_A-\bm{R}_B|^2/\beta^2}
\end{align}
$\bm R^{0}$ is a local average position of the nuclei and $\bm K_B$ is a moment-of-inertial like tensor. The expressions are given in the SI section S2. In the non-local limit, $\zeta_{AB} \rightarrow M_{A}$, $\bm R^{0}$ becomes the center of mass and $\bm K_B$ becomes the nuclear moment of inertia relative to the center of mass. \cite{tao_basis-free_2025}

Although the exact symmetry constraints above do not uniquely define the form of $\hat{\bm \Gamma}$, we have demonstrated previously that the choice of $\hat{\bm \Gamma}$  in Eqs.\ref{eq:gammasum} -\ref{eq:erf} is reasonable by benchmarking against several quantities, such as the electronic momentum, electronic current densities, and vibrational circular dichroism.\cite{duston_phase-space_2024,tao_basis-free_2025,tao_electronic_2024} 
At this point, however, it is crucial to note that, for evaluating Eq.\ref{eq:CIDm_PS} above, we must work {\em in the presence} of a magnetic field.
To compute the ROA tensors in Eq. \ref{eq:CIDm_PS}, we need to explicitly take into account the effects of an external magnetic field. 

\subsubsection{Phase Space Electronic Structure Theory in the Presence of External Magnetic Fields}

Before we address the proper theory of phase space electronic structure in the presence of a uniform magnetic field, let us remind the reader extremely briefly about dynamics in the presence of a $\bm B$ field.  In the presence of a magnetic field, both the nuclei and electrons experience the magnetic field.  
 Within the BO framework, the total Hamiltonian can be separated  naturally as follows:
\begin{align}
\hH^{mag}_{\rm BO}(\bm{R},\bB) &= \sum_{A}\frac{ \bm{\Pi}^2_A(\bm B) }{2M_A} +  \hH^{mag}_{el}(\bm{R}, \bm B),\label{eq:Hmag_bo} \\ 
    \hH^{mag}_{el}(\bm{R}, \bm B) &= \sum_{i}\frac{\hat{\pi}^2_i}{2} + \hat{V}_{eN} + \hat{V}_{ee} + \hat{V}_{NN} \label{eq:Hmag_el}
\end{align}
Here the nuclear and electronic kinetic  (as oppposed to canonical) momentum are 
\begin{align}
    \bm{\Pi}_A &= \bm P_A - Q_A e\bm A(\bm R_A) \\
    \hat{\bm{\pi}}_i &=  \hat{\bm p}_i + e\bm A(\hat{\bm r}_i)
\end{align}
 Above, $\bm A(\bm R) = \frac{1}{2}\bm B\times(\bm R-\bm G_0)$ is the vector potential of the external magnetic field $\bm B$ at position $\bm R$ with a gauge origin at $\bm G_0$. Two points are worth noting at this juncture: (i) The magnetic field in the total Hamiltonian appears only in the kinetic momentum, which makes the entire theory gauge invariant (i.e., all dynamics will be invariant to the choice of $\bm G_0$). (ii)  As pointed out in Refs.\citenum{cederbaum1988,schmelcher_validity_1988,yin1994magnetic}, if one runs dynamics along an eigensurface of 
 Eq. \ref{eq:Hmag_el}, unfortunately, the resulting description lacks electronic shielding; nuclei experience the external magnetic fields as bare nuclear charges (whereas in truth, nuclei always drag an electronic cloud). To capture the missing electronic shielding, it is well-understood that the Berry connection or the diagonal component of the derivative coupling vector has to be included.
\begin{align}
    \hH^{mag}_{\rm ad}(\bm{R},\bB) &= \sum_{A}\frac{ (\bm{\Pi}_A(\bm B) - i\hbar \hat{\bm d}^{A}_{KK} )^2 }{2M_A} +  E^{mag}_{el,KK}(\bm{R}, \bm B)\label{eq:HBO_mag}
\end{align}
This Berry connection term gives rise to a Berry force in addition to the bare nuclear Lorentz force. \cite{resta_manifestations_2000,ceresoli_electron-corrected_2007,culpitt_ab_2021}

From a physically motivated point of view, a phase-space theory of coupled nuclear-electronic motion  must at a minimum include such electronic screening effects; more generally, one would also expect an improved electronic structure wavefunction that  accounts for the electronic motion dragged by the nuclear motion locally in an external magnetic field. 
Although progress on phase space electronic structure in the presence of a magnetic field is quite a bit further behind progress without such a magnetic field, we were able to construct one such approach
in our recently proposed electronic phase-space Hamiltonian in an external magnetic field.\cite{bhati_phase-space_2025-1,bhati_phase-space_2025} Our ansatz was of the form:
    \begin{align}
\hat{H}^{mag}_{PS}(\bm R,\bP,\bm G, \bm B)=\sum_{A}\frac{1}{2M_A} \left( \bm{\Pi}^{\textit{eff}}_A-i\hbar \hbm{\Gamma}_A(\bR) \right)^2+  \hH^{mag}_{el}(\bm R,\bm G, \bm B, \bm F)\label{eq:PSH_mag}
\end{align}
Here, the kinetic nuclear momentum $\bm M \dot{\bm R} = \bm{\Pi}^{\textit{eff}}_A-i\hbar \hbm{\Gamma}_A(\bR)$  
takes into account the static shielding of the nuclear charges experienced by the external magnetic field:
\begin{eqnarray}
\bm{\Pi}^{\textit{eff}}_A&=&\bP_A - \frac{1}{2} q^{\textit{eff}}_A(\bm R)(\bB\times (\bm R_A -\bG_0)) \label{eq:pieffdefn}
\end{eqnarray}
where
$q^{\textit{eff}}_A$ is an effective nuclear charge
\begin{eqnarray}
    q^{\textit{eff}}_A(\bm R) &=&   Q_Ae -e\left< \psi_0\middle|\hat{\Theta}_A(\hbr,\bm R)\middle| \psi_0 \right>\label{eq:qeffdefn}
\end{eqnarray}
Here $\ket{\psi_0}$ here represents the ground state electronic wavefunction with $\bm B= 0$ and $\bm P= 0$. Moreover, the presence of the $\hat{\bm \Gamma}$ operator takes into account the electronic motion that is induced by nuclear motion, which reduces to the standard PS approach in Eq. \ref{eq:PSH} (in the absence of a magnetic field) and  gives a correction to the nuclear Lorentz force similar to a Berry force (in the presence of a magnetic field).

Now, to design a reasonable one-electron $\hat{\bm \Gamma}$ operator, for the most general possible approach, we must revisit the symmetry constraints  in the presence of an external magnetic field. Specifically,
for a charge neutral system, there are no longer six conserved quantities; instead, there are four.
Namely, although the three components of the total canonical momentum $\bm P_{tot} =\sum_A \bm P_{A} + \langle \hat{\bm p}\rangle$ are not conserved, one does conserve the three components  of pseudomomentum $\bm K_{tot} =\sum_A \bm K_{A} + \langle \hat{\bm k}\rangle$. Here, we define:
\begin{align}
    K_{A\alpha} &= P_{A\alpha} + \frac{Q_A e}{2} (\bB \times (\hat \bR_{A}-\bG_0))_\alpha \\
    \hat k_\alpha &=\hat p_\alpha - \frac{e}{2} (\bB \times (\hbr-\bG_0))_\alpha
\end{align}
Furthermore, the total canonical angular momentum in the direction of the external magnetic field is conserved (which is the fourth conserved quantity).
At this point, note that accounting {\em exactly} for these four  conserved quantities above is impossible within a meaningful semiclassical phase space theory because the pseudomomentum is not gauge invariant (unlike the kinetic momentum).  
Nevertheless, to account approximately for the relevant conserved quantities, we have previously proposed the following  constraints for $\hat{\bm \Gamma}$ in an external magnetic field aligned along the z-axis.\cite{bhati_phase-space_2025-1}
\begin{align}
    -i\hbar\sum_{A}{ \hat{\bm \Gamma}}_{A} + \sum_{A}\hat{\Theta}_A{ \hat{\bm k}_A} &= 0,\label{eq:Gamma_uv1} \\
    \label{eq:Gamma_uv2}
    \Big[-i\hbar\sum_{B}\pp{}{{R}_{B\alpha}} + \hat{p}_\alpha -i \hbar \frac{\partial}{\partial G^{0}_\alpha} , \hat{ \Gamma}_{A\beta}\Big] &=0 \\
    -i\hbar\sum_{A}({\bm R}_{A}-\bm G_0) \times {\hat{\bm \Gamma}}_{A} + \sum_{A}\hat{\Theta}_A (\hat{\bm{r}}-\bm{G}_0)\times {\hat{\bm k}_A} &= 0\label{eq:Gamma_uv3}\\
    \Big[-i\hbar\sum_{B}\left((\bm{R}_B-\bG_0) \times\pp{}{\bm{R}_B}\right)_{z} + \left((\hbm{r} - 
    \bG_0) \times \hbm{p}\right)_{z} + \hat{s}_{z}, \hat{\Gamma}_{A \delta}\Big] 
     &= i\hbar \sum_{\alpha} \epsilon_{\alpha z \delta} \hat{\Gamma}_{A \alpha}
     \label{eq:Gamma_uv4}
 \end{align} 
Here $\hat{\bm k}_A$ is the electronic pseudomomentum relative to a nuclear coordinate $\tilde{\bm R}_A$ that moves with the atoms (so that $\hat{\bm{k}}_A$ is translationally invariant):
\begin{eqnarray}
    \hat{\bm{k}}_A  &\equiv& \hat{\bm{k}} +  e\bm{B}\times ( \tilde{\bm R}_A-\bG_0) 
\label{eq:pseudoAdefn}
\end{eqnarray}
In Refs. \citenum{bhati_phase-space_2025-1,bhati_phase-space_2025}, we fixed $\tilde{\bm R}_A$ as the nuclear coordinate $\bm R_A$,
\begin{eqnarray}
    \tilde{\bm R}_A= \bm R_A,  \label{eq:X_A}
\end{eqnarray}
in order to interpret $\hat{\bm k}_A$ as the electronic pseudomomentum partitioned locally to each nuclei. That being said, such a definition was not unique.  Thus, in what follows we will also explore an alternative  definition   of $\hat{\bm k}_A$ as  the electronic pseudomomentum relative to a local nuclear center of charge 
\begin{align}
    \bm R^{A}_{cc}  = \frac{ \sum_B \xi_{AB} Q_B \bm{R}_B}{\sum_B \xi_{AB}Q_B } \label{eq:X_Acc}
\end{align}
or a local center of mass 
\begin{align}
    \bm R^A_{cm} = \frac{ \sum_B \xi_{AB} M_B \bm{R}_B}{\sum_B \zeta_{AB}M_B }\label{eq:X_Acm}
\end{align}
Here, $\xi_{AB}$ is a locality function
\begin{align}
\xi_{AB} =  e^{-|\bm{R}_A-\bm{R}_B|^2/\eta^2}
\label{eq:xi}
\end{align} with a locality parameter $\eta$. For the calculations presented in this paper, we fixed $\eta=\sigma$, although there is some room to choose $\eta$ as long as it is not too local or too non-local. 

If we make such a definition, in order to maintain the proper conservation laws, we must revisit the definitions of $\bm{\Pi}^{\textit{eff}}$ and $\bm q^{\textit{eff}}$ from Eqs. \ref{eq:pieffdefn} and \ref{eq:qeffdefn} above, and set the corresponding effective nuclear charges $ q^{\textit{eff}}_A$ as:
\begin{align}
\bm{\Pi}^{\textit{eff}}_A&=\bP_A - \frac{1}{2} q^{\textit{eff,cc}}_A(\bm R)(\bB\times (\bm R_A -\bG_0))  \\
    q^{\textit{eff,cc}}_A(\bm R) =&   Q_Ae -e\sum_B\left< \psi_0 \middle|\hat{\Theta}_B(\hbr,\bm R)\middle| \psi_0 \right>\frac{  \xi_{AB} Q_A }{\sum_C \xi_{BC}Q_C }\label{eq:qeff_coc}
\end{align}
    or
\begin{align}
   \bm{\Pi}^{\textit{eff}}_A&=\bP_A - \frac{1}{2} q^{\textit{eff,cm}}_A(\bm R)(\bB\times (\bm R_A -\bG_0)) \\
    q^{\textit{eff,cm}}_A(\bm R) =&   Q_Ae -e\sum_B\left< \psi_0 \middle|\hat{\Theta}_B(\hbr,\bm R)\middle| \psi_0 \right>\frac{  \xi_{AB} M_A }{\sum_C \xi_{BC}M_C }\label{eq:qeff_com}
\end{align}
In Fig. S1, we show that the local center-of-charge choice of the electronic psuedomomentum in Eq. \ref{eq:X_Acc} gives better results in calculating the ROA signals than fixing the pseudomomentum relative to each atom (Eq. \ref{eq:X_A}).
Note that, in the non-local limit $\eta\rightarrow \infty $, $\xi_{AB}\rightarrow 1 $, $\bm R^A_{cc}$ and $\bm R^A_{cm}$ in Eqs. \ref{eq:X_Acc}  and \ref{eq:X_Acm}  become the nuclear center of charge and center of mass, respectively.
\begin{eqnarray}
    \hat{\bm{k}}^{A}_{cc}  &\equiv& \hat{\bm{k}} +  e\bm{B}\times ({\bm R^{A}_{cc}}-\bG_0)  \\ \hat{\bm{k}}^{A}_{cm}  &\equiv& \hat{\bm{k}} +  e\bm{B}\times ({\bm R^{A}_{cm}}-\bG_0)
\end{eqnarray}
For a diatomic molecule, all of our experience suggests that the optimal choice is to fix the pseudomomentum ($\hat{\bm k}_A$ in Eq. \ref{eq:pseudoAdefn}) relative  to the  center of mass or charge.

As shown in Ref. \citenum{bhati_phase-space_2025-1}, the symmetry constraints in Eqs.\ref{eq:Gamma_uv1} -\ref{eq:Gamma_uv4} 
are enough to guarantee  
conservation of energy, the total pseudomomentum, and the total canonical angular momentum (in the $z-$direction).
 In SI section S1, we show that changing the reference frame for electronic pseudomomentum to a local nuclear center of charge (Eq. \ref{eq:X_Acc}) or center of mass (Eq. \ref{eq:X_Acm}) does not affect these conservation laws.
To satisfy these symmetry constraints, we will construct a $\hat{\bm \Gamma}$ operator of the following form:\cite{bhati_phase-space_2025-1}
\begin{eqnarray}
   {\hat{ \bm\Gamma}}_A &=&{\hat{ \bm\Gamma}}_A' +{\hat{ \bm\Gamma}}_A''\label{eq:gammasum_mag}\\
    {\hat{ \bm\Gamma}}_A' &=& \frac{1}{2i\hbar}\left( \hat{\Theta}_A(\hbr){\hat{\bm{ k}}_A}+{\hat{\bm{ k}}_A}\hat{\Theta}_A(\hbr)\right)\label{eq:etf_mag}\\
     \hat{\bm{\Gamma}}_A^{''} &=& \sum_{B}  \zeta_{AB}\left(\bm R_A -\bm R^0_{B}\right)\times \left(\bm{K}_B^{-1}\frac{1}{2i\hbar}(\hbm{r}-\bm R_B)\times(\hat{\Theta}_B(\hbr) \hat{\bm{k}}_{B} + \hat{\bm{ k}}_{B}\hat{\Theta}_B(\hbr))\right)\label{eq:erf_mag}
\end{eqnarray}
Here the locality functions $\hat{\Theta}_A(\hbr)$ and $\zeta_{AB}$ are unchanged relative to the case without the external magnetic field.

\subsection{Raman Optical Activity Tensor and a Phase Space Evaluation of W (Eq. \ref{eq:W})} 

\label{sec:PS_W}

It must be emphasized that, until now,  the phase space theory proposed above (in Eq. \ref{eq:PSH_mag} and Eqs. \ref{eq:gammasum_mag}-\ref{eq:erf_mag})  has not been tested against experiment. Indeed, future work should certainly test such a theory against well characterized magnetic field dependent quantities, e.g. the g-factor\cite{bak_vibrational_2005,ekstrom_computational_2025,resta_molecular_2023}. That being said, the goal of this paper is to benchmark such an approach against experimental ROA data. Specifically we wish to evaluate the tensor (see Eq. \ref{eq:dGdp_diff})

\begin{align}
\label{eqn:dWdP:again}
   \frac{\partial W_{\gamma\alpha}}{\partial \bm P_{\beta}} \equiv & \frac{\partial^2 \bra{\psi_{PS}}\hat{ \mu}_{\gamma} \ket{\psi_{PS}}}{\partial B_{\alpha}\partial \bm P_{\beta}}  - \frac{\partial^2 \bra{\psi_{PS}}\hat{ m}_{\alpha} \ket{\psi_{PS}}}{\partial F_{\gamma}\partial \bm P_{\beta}} 
\end{align}
within the electronic phase space theory. The ROA invariants $\alpha G'$ and $\gamma^2$ in Eqs. \ref{eq:ag} and \ref{eq:g2} for the $k$th vibrational mode can then be computed accordingly using Eq. \ref{eq:new0},
\begin{align}
    \omega^{-1}\alpha^{(k)} G'^{(k)} &= \frac{1}{9\omega} \sum_{\alpha\beta} \alpha^{(k)}_{\beta\beta} G'^{(k)}_{\alpha\alpha}\\
    & = \frac{\hbar}{18\omega_k}  \sum_{\alpha\beta A\gamma B\eta} \frac{\partial^{2} \mu_{\beta}}{\partial R_{A\gamma}\partial F_{\beta}} S^{(k)}_{A\gamma}\frac{\partial W_{\alpha\alpha}}{\partial  P_{B\eta}}M_B S^{(k)}_{B\eta}
    \label{eqn:gp_again}
\end{align}
and  
\begin{align}
\omega^{-1}\gamma^{2(k)} & = \frac{1}{2\omega}\sum_{ji} (3\alpha^{(k)}_{\beta\alpha}G'^{(k)}_{\beta\alpha} - \alpha_{\beta\beta}G'^{(k)}_{\alpha\alpha})\\
& = \frac{\hbar}{4\omega_k} \sum_{\alpha\beta A\gamma B\eta} S^{(k)}_{A\gamma}\Bigg[3\frac{\partial^{2} \mu_{\alpha}}{\partial R_{A\gamma}\partial F_{\beta}}\cdot \frac{\partial W_{\beta\alpha}}{\partial  P_{B\eta}} -\frac{\partial^{2} \mu_{\beta}}{\partial R_{A\gamma}\partial F_{\beta}}\frac{\partial W_{\alpha\alpha}}{\partial  P_{B\eta}}\Bigg]M_BS^{(k)}_{B\eta} 
\label{eqn:gam2_again}
\end{align}
At this point, however, we note that the ROA invariants $\alpha G'$ and $\gamma^2$ in Eqs. \ref{eqn:gp_again} and \ref{eqn:gam2_again} are not directly invariant to the gauge origin with PS theory -- which is a big problem.  Most directly, if we plug in Eq. \ref{eqn:dWdP:again} for $\partial \bm W/\partial \bm P$, Eqs. \ref{eqn:gp_again} and \ref{eqn:gam2_again}  will be invariant to origin only if Eq. \ref{eqn:nafie} is obeyed (which is untrue, see Eq. \ref{eqn:no_nafie}). From a fundamental point of view,  Eq. \ref{eqn:rp_same} no longer holds because  $\left[\hat{\bm \Gamma},\hat{\br} \right]\ne 0$, which in turn invalidates Eq. \ref{eqn:nafie} according to the derivation in Ref. \cite{tao_practical_2024}.

The description above necessitates that we use a  distributed origin (DO) scheme \cite{Nafie_2011_VCDbook, duston_phase-space_2024} that evaluates the   term $\frac{\partial \bm m}{\partial \bm P}$ in a  manner that avoids gauge origin dependence.  The ansatz is simple: from the definition of $\hat{\bm m}$ in Eq. \ref{eq:m1} above, we define the ground state AAT within the DO scheme as :
\begin{align}
\label{eq:do:direct}
\Big(\frac{\partial  m_{\alpha}}{\partial P_{B\beta}}\Big)^{DO} & \equiv -  \frac{e }{2m_e}  \frac{\partial  }{\partial P_{B\beta}} \langle (
\hat{\bm r}-\bm R_B) \times \hat{\bm p} \rangle_{\alpha}\\
    &= \frac{\partial m_{\alpha}}{\partial  P_{B\beta}} -  \frac{e }{2m_e} \sum_{\gamma\eta}\epsilon_{\alpha\gamma\eta}(R_{B\gamma}-G^{\gamma}_{0} ) \frac{\partial p_{\eta}}{\partial  P_{B\beta}} \label{eq:dmdP_dpdP}
\end{align}
Note for the phase space electronic dipole moments below, we use the notations above and below $m_{\alpha} = \bra{\psi_{PS}} \hat{m}_{\alpha} \ket{\psi_{PS}}$, $p_{\alpha} = \bra{\psi_{PS}} \hat{p}_{\alpha} \ket{\psi_{PS}}$, and $\mu_{\alpha} = \bra{\psi_{PS}} \hat{\mu}_{\alpha} \ket{\psi_{PS}}$ for simplicity.

Using Nafie's relationship (Eq. \ref{eqn:nafie} above), this above equation can be inverted, which allows us to approximate:
\begin{align}
   \frac{\partial  m_{\alpha}}{\partial P_{B\beta}}
   &\approx 
   \Big( \frac{\partial  m_{\alpha}}{\partial P_{B\beta}} \Big)^{DO}   - \frac{1 }{2M_B} \sum_{\gamma\eta}\epsilon_{\alpha\gamma\eta}(R_{B\gamma}-G^{\gamma}_{0} )\frac{\partial \mu_{\eta}}{\partial  R_{B\beta}}    \label{eq:dmdP_drdR_opposite}
\end{align} 
Eq. \ref{eq:dmdP_drdR_opposite} is exact for Shenvi's Hamiltonian in the limit of a complete basis set; for our approximate phase space Hamiltonian, this statement is not exact, but will form a very useful expression for reasons that will be apparent. 
Note that the ansatz above leads a natural DO version of $\partial \bm W/ \partial \bm P$ :
\begin{eqnarray}
\label{eq:dwdp:do}
   \left( \frac{\partial W_{\gamma\alpha}}{\partial  P_{B\beta}} \right)^{DO} &\equiv&  \frac{\partial^2 \bra{\psi_{PS}}\hat{ \mu}_{\gamma} \ket{\psi_{PS}}}{\partial B_{\alpha}\partial  P_{B\beta}}  -
   \frac{\partial}{\partial F_{\gamma}}
   \left(
   \frac{\partial \bra{\psi_{PS}}\hat{ m}_{\alpha} \ket{\psi_{PS}}}{\partial P_{B\beta}}\right)^{DO} \\
   & \approx &   \left( \frac{\partial W_{\gamma\alpha}}{\partial  P_{B\beta}} \right) + 
\frac{\partial}{\partial F_{\gamma}}
 \left(\frac{1 }{2M_B} \sum_{\sigma\eta}\epsilon_{\alpha\sigma\eta}(R_{B\sigma}-G^{\sigma}_{0} )\frac{\partial \mu_{\eta}}{\partial  R_{B\beta}} \right)
\end{eqnarray}
or equivalently
\begin{eqnarray} 
\label{eq:approx:dwdp}
 \left( \frac{\partial W_{\gamma\alpha}}{\partial  P_{B\beta}} \right) 
 & \approx &
 \left( \frac{\partial W_{\gamma\alpha}}{\partial  P_{B\beta}} \right)^{DO}
 - 
\frac{\partial}{\partial F_{\gamma}}
 \left(\frac{1 }{2M_B} \sum_{\sigma\eta}\epsilon_{\alpha\sigma\eta}(R_{B\sigma}-G^{\sigma}_{0} )\frac{\partial \mu_{\eta}}{\partial  R_{B\beta}} \right)
\end{eqnarray}

Finally, according to Eq.\ref{eq:new1} and Eq. \ref{eq:CIDm_PS} ,  if we plug in Eq. \ref{eq:approx:dwdp}, we find our final approximation for the  key factor in the magic angle CID:
\begin{align}
 \sum_{\alpha\beta }  \frac{\partial \alpha_{\alpha\beta}}{\partial X_{k}} \frac{\partial   W_{\alpha \beta} }{\partial{ Z}_{k}} 
 & =\sum_{\alpha\beta }  \frac{\partial \alpha_{\alpha\beta}}{\partial X_{k}} \Big( \frac{\partial   W_{\alpha \beta} }{\partial{ Z}_{k}}\Big)^{DO} - \frac{1}{2}\sum_{\alpha\beta A\gamma } \frac{\partial^{2} \mu_{\beta}}{\partial R_{A\gamma}\partial F_{\alpha}} S^{(k)}_{A\gamma}\sum_{\delta\sigma B\eta}\Big( \epsilon_{\beta\delta\sigma}R_{B\delta} \frac{\partial^2  \mu_{\sigma}}{\partial F_{\alpha}\partial  R_{B\eta}}\Big) S^{(k)}_{B\eta} \label{eq:CIDm_do}
\end{align}

The  first term  in Eq. \ref{eq:CIDm_do} is origin independent by definition. The second term in Eq. \ref{eq:CIDm_do} is also gauge origin independent-- if one translates the gauge origin by $\bDelta$ (e.g. $\bDelta = \bG_0$), the additional $\Delta$-dependent term arising from the second term in Eq. \ref{eq:CIDm_do} vanishes as 
\begin{align}
    - \frac{1}{2}\sum_{\alpha\beta A\gamma  } \frac{\partial^{2} \mu_{\beta}}{\partial R_{A\gamma}\partial F_{\alpha}} S^{(k)}_{A\gamma}\sum_{\delta\sigma B\eta}\Big( \epsilon_{\beta\delta\sigma}\Delta_{\delta} \frac{\partial^2  \mu_{\sigma}}{\partial F_{\alpha}\partial  R_{B\eta}}\Big) S^{(k)}_{B\eta} = 0 \label{eq:CIDm_go_test}
\end{align}
Next,  we examine the $\alpha G'$ tensor, which is present in the polarized and depolarized CID expressions.  In line with the argument above, we define a DO expression (in tandem with Eq. \ref{eqn:gp_again} and Eq. \ref{eq:dwdp:do})  as the levi-civita symbol $\epsilon_{\beta\delta\sigma}$ is antisymmetric and $\sum_{\alpha}\frac{\partial\bm \mu}{\partial F_{\alpha}}\times\frac{\partial\bm \mu}{\partial F_{\alpha}} = 0$ :
\begin{align}
    & \omega^{-1} \left(\alpha^{(k)} G'^{(k)}\right)^{DO} \nonumber
    \\
    =& \frac{1}{9\omega}  \sum_{\alpha\beta} \alpha_{\beta\beta}^{(k)} \left(G'^{(k)}_{\alpha\alpha}\right)^{DO}\label{eq:early:aG_do0}\\     
     = &  \frac{\hbar}{18\omega_k}\sum_{\alpha\beta A\gamma B\eta} \frac{\partial^{2} \mu_{\beta}}{\partial R_{A\gamma}\partial F_{\beta}} S^{(k)}_{A\gamma}\Big(\frac{\partial^{2} \mu_{\alpha}}{\partial P_{B\eta}\partial B_{\alpha}} - 
    \frac{\partial}{\partial F_{\alpha}}
    \left(
    \frac{\partial m_{\alpha}}{\partial P_{B\eta}} \right)^{DO} \Big)M_B S^{(k)}_{B\eta}   
    \label{eq:early:aG_do1}
\end{align}
and then we may approximate:
\begin{align}
     \omega^{-1} \left(\alpha^{(k)} G'^{(k)}\right)
    & \approx      \omega^{-1} \left(\alpha^{(k)} G'^{(k)}\right)^{DO} - \frac{\hbar}{36\omega_k} \sum_{\beta A\gamma} \frac{\partial^{2} \mu_{\beta}}{\partial R_{A\gamma}\partial F_{\beta}} S^{(k)}_{A\gamma}\Big(\sum_{\alpha\delta\sigma B\eta} \epsilon_{\alpha\delta\sigma}R_{B\delta} \frac{\partial^2  \mu_{\sigma}}{\partial F_{\alpha}\partial  R_{B\eta}}S^{(k)}_{B\eta} \Big)  \label{eq:aG_do}
\end{align}
In 
 Eq. \ref{eq:aG_do} above, the first  term is automatically gauge origin independent.  For the second term, if one translates the gauge origin by $\bm \Delta$ (e.g. $\bDelta = \bG_0$), the additional $\bm \Delta$-dependent term  also vanishes:
\begin{align}
    & - \frac{\hbar}{36\omega_k} \sum_{\beta A\gamma} \frac{\partial^{2} \mu_{\beta}}{\partial R_{A\gamma}\partial F_{\beta}} S^{(k)}_{A\gamma}\Big(\sum_{\alpha\delta\sigma B\eta} \epsilon_{\alpha\delta\sigma}\Delta_{\delta} \frac{\partial^2  \mu_{\sigma}}{\partial F_{\alpha}\partial  R_{B\eta}}S^{(k)}_{B\eta} \Big)\nonumber\\
     =&- \frac{\hbar}{36\omega_k} \sum_{\beta A\gamma} \frac{\partial^{2} \mu_{\beta}}{\partial R_{A\gamma}\partial F_{\beta}} S^{(k)}_{A\gamma}\Big[\sum_{\delta B\eta} \Delta_{\delta} \Big(\sum_{\alpha\sigma}\epsilon_{\alpha\delta\sigma}\frac{\partial^2  \mu_{\sigma}}{\partial F_{\alpha}\partial  R_{B\eta}}\Big)S^{(k)}_{B\eta} \Big]=0\label{eq:aG_do2}
\end{align}
The  bracketed term in Eq. \ref{eq:aG_do2} is zero because  the polarizability $\alpha_{\sigma\alpha}=\partial  \mu_{\sigma}/\partial F_{\alpha}$ is symmetric upon exchange of indices $\sigma$ and $\alpha$   whereas the levi-civita symbol $\epsilon_{\alpha\delta\sigma}$ is antisymmetric upon exchange of indices.

For the magic angle case, Eqs. \ref{eq:CIDm_do}, \ref{eq:dwdp:do}, \ref{eq:do:direct} and \ref{eq:CIDm_PS} are the final equations; the final equation is invariant to magnetic origin. As shown in Appendix \ref{sec_dHPS}, the expression for $\frac{\partial W_{\gamma\alpha}}{\partial \bm P_{\beta}}$ in Eq.\ref{eqn:dWdP:again}  can be recast in terms of energy gradients, 
\begin{align}
    \frac{\partial W_{\alpha\beta}}{\partial \bm P_{A\eta}} = \frac{i\hbar}{ M_{A}}  \frac{\partial}{\partial F_{\alpha}} \left\langle \psi_{PS} \middle| \frac{\partial \hat{\bm\Gamma}_{A \eta}}{\partial  B_{\beta}} \middle| \psi_{PS} \right\rangle + i\hbar\frac{\partial^2}{\partial F_{\alpha} \partial  P_{A\eta}}\left\langle \psi_{PS} \middle|   \frac{\partial \bm{\Pi}^{\textit{eff}}}{\partial  B_{\beta}}   \cdot\frac{\hat{\bm\Gamma}}{\bm M}\middle| \psi_{PS} \right\rangle. \label{eq:dwdP_co}
\end{align}
Using the distributed gauge scheme described above in Eq. \ref{eq:CIDm_do}, the final key component in the magic-angle CID expression becomes: 
\begin{eqnarray}
\label{eq:final_original_do}
 \sum_{\alpha\beta }  \frac{\partial \alpha_{\alpha\beta}}{\partial X_{k}} \frac{\partial   W_{\alpha \beta} }{\partial{ Z}_{k}} & = & \sum_{\alpha \beta A\eta}    \frac{\partial \alpha_{\alpha\beta}}{\partial X_{k}}  \Bigg( i\hbar \frac{\partial}{\partial F_{\alpha}} \left\langle \psi_{PS} \middle| \frac{\partial \hat{\bm\Gamma}_{A \eta}}{\partial  B_{\beta}} \middle| \psi_{PS} \right\rangle + i\hbar\frac{\partial^2}{\partial F_{\alpha} \partial  P_{A\eta}}\left\langle \psi_{PS} \middle|   \frac{\partial \bm{\Pi}^{\textit{eff}}\cdot \hat{\bm\Gamma}}{\partial  B_{\beta}}  \middle| \psi_{PS} \right\rangle
 \nonumber
\\ & &  +  \frac{1 }{2}  
 \sum_{\gamma\sigma}\epsilon_{\alpha\gamma\sigma}(R_{A\gamma} - G^{\gamma}_0)
 \left(  \frac{eM_A}{m_e}
 \frac{\partial^2 p_{\sigma}}{\partial  P_{A\eta} \partial F_{\alpha}} 
 +
 \frac{\partial^2  \mu_{\sigma}}{\partial F_{\alpha}\partial  R_{A\eta}}
 \right)
\Bigg) S^{(k)}_{A \eta} 
\end{eqnarray}

For the polarized and depolarized cases, the final equations are given in Table \ref{tab:CID} (where we must replace $\bm G'$ with $\bm G'^{(k)}$, etc) and Eqs. \ref{eq:aG_do}, \ref{eq:early:aG_do1}, and \ref{eq:do:direct}.  To evaluate $\alpha^{(k)} G'^{(k)}$, the relevant expression of interest is obviously:
\begin{eqnarray}
\label{eq:final_original2_do}
 \sum_{\alpha\beta }  \frac{\partial \alpha_{\alpha\alpha}}{\partial X_{k}} \frac{\partial   W_{\beta \beta} }{\partial{ Z}_{k}} & = & \sum_{\alpha \beta A\eta}    \frac{\partial \alpha_{\alpha\alpha}}{\partial X_{k}}  \Bigg(i\hbar  \frac{\partial}{\partial F_{\beta}} \left\langle \psi_{PS} \middle| \frac{\partial \hat{\bm\Gamma}_{A \eta}}{\partial  B_{\beta}} \middle| \psi_{PS} \right\rangle + i\hbar\frac{\partial^2}{\partial F_{\beta} \partial  P_{A\eta}}\left\langle \psi_{PS} \middle|   \frac{\partial \bm{\Pi}^{\textit{eff}}\cdot \hat{\bm\Gamma}}{\partial  B_{\beta}}  \middle| \psi_{PS} \right\rangle
 \nonumber
\\ & &  +  \frac{1 }{2}  
 \sum_{\gamma\sigma}\epsilon_{\beta\gamma\sigma}(R_{A\gamma} - G^{\gamma}_0)
 \left(  \frac{eM_A}{m_e}
\frac{\partial^2 p_{\sigma}}{\partial  P_{A\eta} \partial F_{\beta}} 
 +
 \frac{\partial^2  \mu_{\sigma}}{\partial F_{\beta}\partial  R_{A\eta}}
 \right)
\Bigg) S^{(k)}_{A \eta} 
\end{eqnarray}
For the latter two cases, all equations are in fact invariant to the choice of the gauge origin (up to the considerations in Appendix \ref{sec_delta_G} for the quadrupole terms).
The invariance in Eqs. \ref{eq:final_original_do} and \ref{eq:final_original2_do} might at first appear confusing insofar as $\bm G_0$ does appear, but note that the first three terms in Eqs. \ref{eq:final_original_do} and \ref{eq:final_original2_do}  arise from a distributed origin calculation (as in Eqs. \ref{eq:early:aG_do0}-\ref{eq:aG_do}); put differently, all G-dependence for the $\bm\Gamma$ terms (as shown in SI section S2) is canceled by the $\frac{\partial^2 p_{\sigma}}{\partial  P_{A\eta} \partial F_{\beta}} $ term. The last terms in Eqs. \ref{eq:final_original_do} and \ref{eq:final_original2_do} do not depend on the choice of the gauge origin as shown in Eq. \ref{eq:CIDm_go_test} and Eq.\ref{eq:aG_do2} above.

\section{Results}
\label{sec:results}
We have tested the phase space approach to ROA (as established above) 
against experimental (R)-methyloxirane data. For these preliminary calculations, we computed optimized geometries, Hessians and related quantities (e.g. frequencies, S-vectors) for (R)-methyloxirane within the Hartree-Fock (HF) level of theory with aug-cc-pVQZ (aQZ) basis set at the zero fields. We then computed the ROA signals, where the electric-dipole magnetic-dipole polarizability $\omega^{-1}\bm G'$ was obtained with the PS approach according to Eq. \ref{eq:new1}  and with the conventional perturbation theory according to Eq. \ref{eq:dGdq}. We implemented these methods in a developmental branch of Q-Chem software package. \cite{Epifanovsky2021} In the BO approach, we computed the ROA tensors with aug-cc-pVTZ basis set and the gauge origin ($\bm G_0$) was taken at the center of charge, coinciding with the coordinate origin.  To calculate $\omega^{-1}\bm G'$ with the PS approach, we evaluated Eq. \ref{eq:approx:dwdp} semi-numerically using the distributed origin scheme. To match with experimental data in Refs. \citenum{bose_ab_1990} and \citenum{polavarapu_vibrational_1993} , we used $\omega = 488nm$ in evaluating CIDs in equations such as Eq. \ref{eq:CIDm_BO} and \ref{eq:CIDm_PS}.

\begin{figure}[!htbp]
    \centering
    \includegraphics[width=0.9\linewidth]{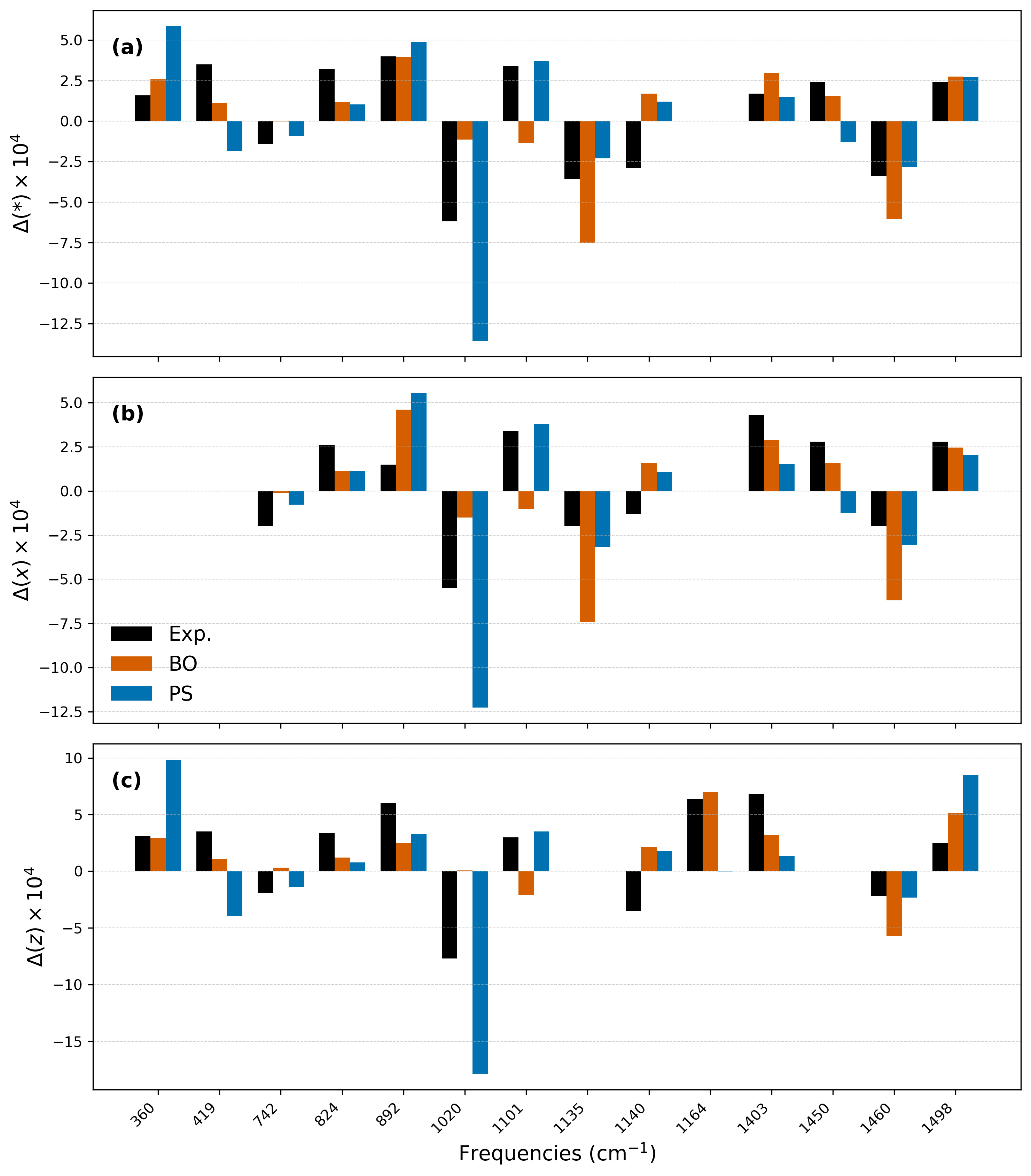}
    \caption{Calculated (a) magic-angle ($\Delta(*)$), (b) polarized ($\Delta(x)$), and depolarized CIDs ($\Delta(z)$) for (R)-methyloxirane using perturbation theory based on BO states (shown in orange) and PS theory (shown in blue, locality parameters, $\sigma=\eta = 1.5$ Bohr and $\beta = 9$ Bohr) compared to experimental values (shown in black). The results shown here were calculated with aug-cc-pVTZ basis set. The magic-angle and polarized experimental results were reported in Ref. \citenum{bose_ab_1990}. For the depolarized experimental results, we used the experimental data from a later experiment in Ref. \citenum{polavarapu_vibrational_1993} that matched the same signs with the previous reported signals in Ref. \citenum{bose_ab_1990}.  Experimentally reported frequencies were used to label x-axis.  }
    \label{fig:do_atz}
\end{figure}

The calculated ROA CIDs with the BO approach and the PS approach are shown in Fig. \ref{fig:do_atz}. Three experimental setups from Ref. \citenum{polavarapu_ab_1990} are considered here: magic-angle, polarized, and depolarized measurements, presented in Fig. \ref{fig:do_atz} (a), (b), and (c), respectively. The ROA CIDs calculated with the conventional perturbation approach using Eq. \ref{eq:CIDm_BO}  and the phase space approach using Eq. \ref{eq:CIDm_PS} are shown in orange and blue, respectively. The experimental values reported from Ref. \citenum{bose_ab_1990} and \citenum{polavarapu_vibrational_1993} are shown in black. Note that for the depolarized experimental CIDs, some signals were reported with inconsistent signs between the two references. We only plotted the ones with consistent signs in Fig.  \ref{fig:do_atz} (c). We label the x-axis with the experimental vibrational frequencies and align the computed CIDs to the same frequency order observed experimentally.
Admittedly, previously studies have demonstrated that frequency orders can change upon using larger basis sets and including electron-electron correlation. \cite{polavarapu_vibrational_1993} To that end, for the frequency calculations, we use a large electronic basis set (aQZ) to minimize the frequency dependence on basis set choice. To investigate the effect of electron-electron correlation on the ordering of the vibrational modes, we computed vibrational frequencies with density functional theory (DFT)/B3LYP (as opposed to HF) and an aQZ basis set  in Table S4. In comparing the vibrational frequencies and modes computed with HF and DFT/B3LYP, we found that the peaks at 1267 cm$^{-1}$ and 1296 cm$^{-1}$ appear in reversed order, corresponding to the experimental peaks at 1135 and 1140 cm$^{-1}$ shown in Fig. \ref{fig:do_atz}. As the experimental signals at these two peaks are very similar, these changes in ordering do not affect the results discussed below.

Comparing the CIDs calculated by the two approaches and also with the experimental CIDs under the three different setups in Fig. \ref{fig:do_atz}, we first observe that the two approaches perform similarly in terms of the number of correct signs predicted. For all the three measurements, both the conventional BO perturbation approach and the PS approach predict an incorrect sign at 1140 cm$^{-1}$, which corresponds to a combination of CH\textsubscript{2} twisting and C*--CH\textsubscript{3} stretching modes, where the asterisk (*) indicates the chiral center.  For both magic-angle and polarized CIDs, the conventional BO perturbation approach further predicts incorrect signs at 1101 cm$^{-1}$ and very weak signals at 742 cm$^{-1}$, which correspond to a combination of CH$_2$ and CH$_3$ rocking modes and a combination of the CH$_3$-C*, O-C*, and C*-CH$_2$ stretching modes, respectively. Similarly, the PS approach predicts incorrect signs at 419 cm$^{-1}$ for the magic-angle CID, and 1450 cm$^{-1}$ for both magic-angle and polarized CIDs, which correspond to CH$_3$-C*-O bending mode  and CH$_3$ asymmetric bending mode. For the depolarized setup, the BO approach additionally gives a weak signal at 1020 cm$^{-1}$, which corresponds to  a combination of CH$_3$ rocking, CH$_2$ twisting, and  C*--H bending mode, respectively. The PS approach gives incorrect signals at 419 cm$^{-1}$ and a weak signal at 1164 cm$^{-1}$, the latter corresponding to a  C*--H bending mode. Overall, both approaches give a reasonable match with experimental data. Further improvements can be made by going beyond the approximations mentioned in the introduction (e.g. including electron-electron correlation, etc).

\section{Discussion: Robustness of the PS approach and Room for Improvement}

Above, we have demonstrated that a PS approach to electronic structure theory in the presence of a magnetic field can be directly tested by comparison with experimental raman optical activity spectra and standard BO Berry curvature approaches.  Our results are  encouraging and invariant to origin (at least without considering the quadrupole term; see Appendix \ref{sec_delta_G}). Our approach is also less expensive than standard BO theory insofar as we do not require a calculation of  the Berry curvature derivative in Eq. \ref{eq:dGdq}; instead, we calculate $\frac{\partial^2\bm m}{\partial \bm P\partial \bm F}$ or  $\frac{\partial^2\bm p}{\partial \bm P\partial \bm F}$, which requires an additional orbital response calculation with respect to the nuclear momentum, but the Fock derivative only contains an one-electronic component.  See Eqs. \ref{eq:final_original_do} and \ref{eq:final_original2_do}.

In the future, as a means of improving the present simulations, we can envision two important directions. First, although not emphasized above, all of our data does depend on the choice of locality parameters $\sigma$ in Eq. \ref{eq:theta} and $\eta$ in Eq. \ref{eq:xi} for the ETF component  and $\beta$ in Eq. \ref{eq:zeta} for the ERF component. For calculations shown in Fig. \ref{fig:do_atz}, we chose $\sigma=\eta=1.5$ Bohr and $\beta= 9$ Bohr.  Thus far, we have built up quite a lot of experience as far as understanding whence these parameters arise. In short, if the locality parameters are too small, the partition between different regions of nuclei give rise to sharp discontinuities; if these values are too large, there is no locality and the calculations give no signal or very weak signals. For example, if one uses prefactors in the locality function as mass ratios, as in Eq. \ref{eq:X_Acm}, at the non-local limit, the electrons are only coupled to the nuclear center of mass, which do not change upon vibrations, and hence would give zero signals.\cite{tao_basis-free_2025} Thus, for PS to be applicable, there must be a robust sweet spot and future work will need to benchmark results against different parameters and assert the robustness of the present ROA approach. So far, despite reasonable accuracy we must admit that our ROA predictions do appear somewhat less robust than VCD data\cite{duston_phase-space_2024}; this observation would appear reasonable since ROA requires one more derivative than VCD and thus should be  more sensitive to the approximations in a phase space approach.

Second,  a few words are appropriate regarding the choice of symmetries that underlie Bhati's approach to magnetic field chemistry (in Refs. \citenum{bhati_phase-space_2025-1,bhati_phase-space_2025}) that has formed the basis for the present study. The theory in Refs. \citenum{bhati_phase-space_2025-1,bhati_phase-space_2025} was based on the notion that we ought to build the $\hat{\bm \Gamma}$ operator so as to conserve the relevant momenta of the matter degrees of freedom, i.e. without explicitly considering the momentum of the external fields. However, one can ask if one should take a further step by also considering the momentum of the external fields. Specifically, if we include the external fields, as contrasted with Eqs. \ref{eq:Gamma_uv1}-\ref{eq:Gamma_uv4} above, one might wonder if one could solve a more general set of constraints for the $\hat{\bm \Gamma}$ operators, e.g. of the form:
\begin{align}
    -i\hbar\sum_{A}{ \hat{\bm \Gamma}}_{A} + \sum_{A}\hat{\Theta}_A{ \hat{\bm k}_A} + \hat{\bm P}_F + \hat{\bm P}_B &= 0,\label{eq:Gamma_fb1} \\
    \label{eq:Gamma_fb2}
    \Big[-i\hbar\sum_{B}\pp{}{{R}_{B\alpha}} + \hat{p}_\alpha -i \hbar \frac{\partial}{\partial G^{0}_\alpha}+ \hat{ P}^{\alpha}_F + \hat{ P}^{\alpha}_B , \hat{ \Gamma}_{A\beta}\Big] &=0 \\
    -i\hbar\sum_{A}({\bm R}_{A}-\bm G_0) \times {\hat{\bm \Gamma}}_{A} + \sum_{A}\hat{\Theta}_A (\hat{\bm{r}}-\bm{G}_0)\times {\hat{\bm k}_A} + \hat{\bm J}_F + \hat{\bm J}_B&= 0\label{eq:Gamma_fb3}\\
    \Big[-i\hbar\sum_{B}\left((\bm{R}_B-\bG_0) \times\pp{}{\bm{R}_B}\right)_{z} + \left((\hbm{r} - 
    \bG_0) \times \hbm{p}\right)_{z} + \hat{s}_{z}+ \hat{ J}^{z}_F + \hat{ J}^{z}_B, \hat{\Gamma}_{A \delta}\Big] 
     &= i\hbar \sum_{\alpha} \epsilon_{\alpha z \delta} \hat{\Gamma}_{A \alpha}
     \label{eq:Gamma_fb4}
 \end{align} 
Here $\hat{\bm P}_F  = -i\hbar \frac{\partial}{\partial \bm F}$, $\hat{\bm P}_B  = -i\hbar\frac{\partial}{\partial \bm B}$, $\hat{\bm J}_F  = -i\hbar\bm F\cross\frac{\partial}{\partial \bm F}$, and $\hat{\bm J}_B  = -i\hbar \bm B\cross\frac{\partial}{\partial \bm B}$. Eq. \ref{eq:Gamma_fb1} and Eq. \ref{eq:Gamma_fb3} take into account of the linear and angular momentum of the moving fields and Eq. \ref{eq:Gamma_fb2} and  Eq. \ref{eq:Gamma_fb4} ensure the translational and rotational invariance of the PS energy.
Alas, we do not know how to find solutions to such equations in any meaningful sense, but we wonder if the curious reader will have better luck than have we.

\section{Conclusions and Future Outlooks}
In this paper, we have presented a new approach to the calculation  of the ROA tensors from the perspective of phase electronic structure theory. Specifically, we have shown that the electric-dipole magnetic-dipole polarizability $\bm G'$ can be viewed as the difference between changes of the electronic transition dipole moments with nuclear momentum in the presence of external electric and magnetic fields (Eqs. \ref{eq:dGdp_diff}-\ref{eq:ps_ROA_tensor}), which in turn can be calculated   within  electronic phase space electronic structure theory (Eq.\ref{eq:approx:dwdp}) using the distributed origin scheme. The ROA CIDs can then be computed within the phase space framework in Eq. \ref{eq:final_original_do} and Eq. \ref{eq:final_original2_do}. Of note, the present approach is  invariant to the choice of gauge origin and should be stable. We have performed calculations  for a prototypical molecule, \textbf{R}-methyloxirane, and found reasonable agreement with the experimental data.   A more systematic parameterization will be important to generalize to other systems and more efficient grid evaluation will be needed to take advantage of the simple one-electron form of the $\hat{\bm \Gamma}$ operator.

Finally, we conclude with a few words about the Berry connection and curvature, which plays a key role in quantum geometry, studying intriguing phenomenon such as the anomalous Hall effect\cite{nagaosa_anomalous_2010} and topological insulators in materials\cite{ando_topological_2013,krishnamoorthy_topological_2023} and which has recently been discussed in the context of chiral phonons\cite{zhang_angular_2014,ren_adiabatic_2024,hernandez_observation_2023}, which is deeply related to the CISS phenomenon\cite{fransson_chiral_2023,li_chiral_2024}. 
The importance of Berry curvature for vibrational circular dichroism has also been pointed out in Ref.\citenum{resta_molecular_2023}.
Despite the many successes of the Berry curvature approach, we must emphasize that in this work, we have avoided calculating the Berry curvature. Indeed, a phase space electronic structure theory approach can calculate dynamical quantities that are not available within BO theory by avoiding the BO framework entirely, and moreover, we find that in principle, a PS can be less expensive than a BO berry curvature -- note that we require one less derivative in Eq. \ref{eq:new1} and Eq. \ref{eq:approx:dwdp} than in Eq. \ref{eq:dGdq}.  Furthermore, in the Appendix \ref{sec_dHPS}, we will show that if we are prepared to sacrifice gauge origin invariance, one can obtain an even cheaper estimate of ROA intensities (and save on two derivatives). As such,  our hope is that the method can be used for a host of other applications as well, including some of the  effects listed above (and more) in the future.

\begin{acknowledgments}
This work was supported by the U.S. Department of Energy,
Office of Science, Office of Basic Energy Sciences, under Award
No. DOE-SC0025393 The authors thank Nadine Bradbury, Titouan Duston, Linqing Peng, and Xuecheng Tao for useful discussion. 
\end{acknowledgments}

\section{Appendix}

\subsection{Gauge Invariance for the Quadrupole Containing Term, $\delta^2$ (Eq. \ref{eq:d2})}\label{sec_delta_G}

In the text above, we have argued that the phase space approach can be used to develop an approach to compute the ROA invariants $\alpha G'$ and $\delta^2$ in a gauge-invariant fashion.  Formally speaking, however, this invariance holds only for the magic-angle case. For the case of polarized and depolarized measurements, however, one encounters the quadrupole terms through the electric-dipole-electric-quadrupole polarizability $\bm A$.   Let us now show that, formally speaking, all terms involving the quadrupole will be gauge origin invariant. Recall the definitions from Eq. \ref{eq:a_real} and \ref{eq:A_real} above.
\begin{align}
    {\alpha}_{ij} & = \frac{2}{\hbar} \sum_{J\ne I} \frac{\omega_{JI}}{\omega^2_{JI} - \omega^2} \text{Re}\Big[ \bra{\Psi_I} \hat{\mu}_{i}\ket{\Psi_J} \bra{\Psi_J} \hat{\mu}_{j}\ket{\Psi_I}\Big]\\
    {A}_{lin} & = \frac{2}{\hbar} \sum_{J\ne I} \frac{\omega_{JI}}{\omega^2_{JI} - \omega^2} \text{Re}\Big[ \bra{\Psi_I} \hat{\mu}_{l}\ket{\Psi_J} \bra{\Psi_J} \hat{\theta}_{in}\ket{\Psi_I}\Big] 
\end{align}
Consider now the dependence of $\bm A$ on the gauge origin $\bG_0$. Using Eq. \ref{eq:theta1}, we can show that:
\begin{eqnarray}
{A}_{lin}(\bG_0) - {A}_{lin}(0) &= &
  \frac{2}{\hbar} \sum_{J\ne I} \frac{\omega_{JI}}{\omega^2_{JI} - \omega^2} \text{Re}\Big[ \bra{\Psi_I} \hat{\mu}_{l}\ket{\Psi_J} \bra{\Psi_J} 
    \frac{3}{2}\hat{\mu}_{i}G^0_{n}
    +\frac{3}{2}G^0_{i}\hat{\mu}_{n}
    + \delta_{in} \sum_{m} \hat{\mu}_{m}G^0_{m} 
    \ket{\Psi_I}\Big] 
    \nonumber \\
    & = & \frac{3}{2} \alpha_{li}G^0_{n}  + \frac{3}{2} \alpha_{ln}G^0_{i}  + \frac{2}{\hbar} \delta_{in} \sum_{m} \alpha_{lm} G^0_{m}
    \label{eq:theta_G}
\end{eqnarray}
Note that terms with the $\bm G_0$-squared dependence in Eq. \ref{eq:theta_G} are zero because they are multiplied with the overlap between two orthogonal states $\bra{\Psi_J}\ket{\Psi_I} = 0$. Let us then evaluate $\delta^2$ as defined in Eq. \ref{eq:d2}:
\begin{eqnarray}
    \frac{\delta^2}{\omega} & = & \frac{1}{2}\sum_{ji} \alpha_{ji}\sum_{ln}\epsilon_{jln}A_{lni},
\end{eqnarray}
We need to show that the following term vanishes:
\begin{eqnarray}
        \sum_{ijln} \alpha_{ji} \epsilon_{jln} \left( \frac{3}{2} \alpha_{li}G^0_{n}  + \frac{3}{2} \alpha_{ln}G^0_{i}  + \frac{2}{\hbar} \delta_{in} \sum_{m} \alpha_{lm} G^0_{m} \right)\label{eq:theta_G2}
\end{eqnarray}
The first term vanishes because $\sum_i \alpha_{ji} \alpha_{li}$ is symmetric in $j$ and $l$, whereas $\epsilon_{jln}$ is antisymmetric. The second term vanishes because $\alpha_{ln}$ is symmetric in $l$ and $n$, whereas $\epsilon_{jln}$ is antisymmetric. The third term vanishes because $\sum_i \alpha_{ji} \delta_{in} = \alpha_{jn}$ is symmetric in $j$ and $n$, whereas $\epsilon_{jln}$ is antisymmetric.

Finally, note that this entire argument holds if we replace $\alpha_{ij}$ with $\alpha_{ij}^{(k)}$, so that the argument is completely robust for a ROA calculation. That being said, it is important to note that this entire argument relies on Eq. \ref{eq:theta_G}, which holds only in an infinite (complete) basis  that satisfies Placzek's identity (i.e. so that Eqs. \ref{eq:theta_G} -\ref{eq:theta_G2}  above hold).   This set of affairs can{\em not} be ameliorated by using GIAOs\cite{helgaker1991giaome}; one simply requires a large basis. Nevertheless, as shown in Fig. \ref{fig:param}, the origin dependence is very weak for these terms for the molecule studied here.

\subsection{Derivation of Eq. \ref{eq:dwdP_co} and Further Approximations}
\label{sec_dHPS}

In this section, we derive  Eq. \ref{eq:dwdP_co}  within phase space electronic structure theory starting from Eq. \ref{eqn:dWdP:again}  To make progress, as noted above (Eqs. \ref{eq:dH_dB}-\ref{eq:dH_dF}),
according to BO theory and the Hamiltonian in Eq. \ref{eq:H}, one can calculate derivatives with respect to the external magnetic field in a straightforward manner using Hellman-Feynman theorem. 
\begin{align}
    \frac{\partial\bra{\psi}\hat{H}\ket{\psi}}{\partial \bm B}= \bra{\psi}\frac{\partial\hat{H}}{\partial \bm B} \ket{\psi} &= -\bra{\psi}\hat{\bm m} \ket{\psi}\\
    \frac{\partial\bra{\psi}\hat{H}\ket{\psi}}{\partial \bm F} = \bra{\psi}\frac{\partial\hat{H}}{\partial \bm F} \ket{\psi}  &= -\bra{\psi}\hat{\bm \mu} \ket{\psi}\label{eq:dHps_dF}
\end{align}

By contrast, however, within an electronic phase space framework, additional derivatives must be taken into account. Namely, note that:
\begin{align}    \frac{\partial\bra{\psi_{PS}}\hat{H}_{PS}\ket{\psi_{PS}}}{\partial \bm B} =& \bra{\psi_{PS}}\frac{\partial\hat{H}_{PS}}{\partial \bm B} \ket{\psi_{PS}} \\
    =& -\bra{\psi_{PS}}\hat{\bm m} \ket{\psi_{PS}} - i\hbar  \frac{\bm{P}}{\bm M} \cdot \left\langle \psi_{PS} \middle| \frac{\partial \hat{\bm\Gamma}}{\partial \bm B} \middle| \psi_{PS} \right\rangle \nonumber\\
    &- i\hbar \frac{\partial \bm{\Pi}^{\textit{eff}}}{\partial \bm B}   \cdot \left\langle \psi_{PS} \middle| \frac{\hat{\bm\Gamma}}{\bm M}\middle| \psi_{PS} \right\rangle\label{eq:dHps_dB} \\
    \frac{\partial\bra{\psi_{PS}}\hat{H}_{PS}\ket{\psi_{PS}}}{\partial \bm F} =& \bra{\psi_{PS}}\frac{\partial\hat{H}_{PS}}{\partial \bm F} \ket{\psi_{PS}}  \\
     =& -\bra{\psi_{PS}}\hat{\bm \mu} \ket{\psi_{PS}} - i\hbar  \frac{\bm{P}}{\bm M}\cdot \left\langle \psi_{PS} \middle| \frac{\partial \hat{\bm\Gamma}}{\partial \bm F} \middle| \psi_{PS} \right\rangle\label{eq:dHps_dE}
\end{align}
where
\begin{align}
    \frac{\partial {\Pi}^{\textit{eff}}_{A\alpha}}{\partial  B_{\beta}} =  - \frac{1}{2}  q^{\textit{eff}}_A  \sum_{\gamma}\epsilon_{\alpha\beta\gamma}(R_{A\gamma} -G^{\gamma}_{0})
\end{align}
Re-arranging terms in Eqs. \ref{eq:dHps_dB} and \ref{eq:dHps_dE}, we find:
\begin{align}
    \bra{\psi_{PS}}\hat{\bm m} \ket{\psi_{PS}} =& -\frac{\partial\bra{\psi_{PS}}\hat{H}_{PS}\ket{\psi_{PS}}}{\partial \bm B} - i\hbar  \frac{\bm{P}}{\bm M} \cdot \left\langle \psi_{PS} \middle| \frac{\partial \hat{\bm\Gamma}}{\partial \bm B} \middle| \psi_{PS} \right\rangle\nonumber\\
    &- i\hbar \frac{\partial \bm{\Pi}^{\textit{eff}}}{\partial \bm B}   \cdot \left\langle \psi_{PS} \middle| \frac{\hat{\bm\Gamma}}{\bm M}\middle| \psi_{PS} \right\rangle\\
    \bra{\psi_{PS}}\hat{\bm \mu} \ket{\psi_{PS}}  =&-\frac{\partial\bra{\psi_{PS}}\hat{H}_{PS}\ket{\psi_{PS}}}{\partial \bm F} - i\hbar  \frac{\bm{P}}{\bm M}\cdot \left\langle \psi_{PS} \middle| \frac{\partial \hat{\bm\Gamma}}{\partial \bm F} \middle| \psi_{PS} \right\rangle
\end{align}
Next we evaluate the derivatives with respect to nuclear momentum and the external fields, which are important for the ROA tensor as shown in Eq. \ref{eq:ps_ROA_tensor}.
\begin{align}
    \frac{\partial^2 \bra{\psi_{PS}}\hat{\bm m} \ket{\psi_{PS}}}{\partial\bm F\partial\bm P} =& -\frac{\partial^3\bra{\psi_{PS}}\hat{H}_{PS}\ket{\psi_{PS}}}{\partial\bm F\partial\bm P\partial \bm B} -  \frac{i\hbar }{\bm M} \frac{\partial}{\partial\bm F} \left\langle \psi_{PS} \middle| \frac{\partial \hat{\bm\Gamma}}{\partial \bm B} \middle| \psi_{PS} \right\rangle\nonumber \\
    &- i\hbar \frac{\partial \bm{\Pi}^{\textit{eff}}}{\partial \bm B}   \cdot \frac{\partial^2}{\partial \bm F \partial \bm P}\left\langle \psi_{PS} \middle| \frac{\hat{\bm\Gamma}}{\bm M}\middle| \psi_{PS} \right\rangle\label{eq:ps_dmdFdP}\\
    \frac{\partial^2 \bra{\psi_{PS}}\hat{\bm \mu} \ket{\psi_{PS}}}{\partial\bm B\partial\bm P}  =&-\frac{\partial^3\bra{\psi_{PS}}\hat{H}_{PS}\ket{\psi_{PS}}}{\partial\bm B\partial\bm P\partial \bm F} -   \frac{i\hbar}{\bm M} \frac{\partial}{\partial\bm B} \left\langle \psi_{PS} \middle| \frac{\partial \hat{\bm\Gamma}}{\partial \bm F} \middle| \psi_{PS} \right\rangle
 \label{eq:ps_dmudBdP}
\end{align}
Taking the difference between Eq.\ref{eq:ps_dmudBdP} and  Eq.\ref{eq:ps_dmdFdP}, the first term on the right hand side of both equations cancel and we are left with
\begin{align}
   \frac{\partial W_{\gamma\alpha}}{\partial \bm P_{\beta}} \equiv & \frac{\partial^2 \bra{\psi_{PS}}\hat{ \mu}_{\gamma} \ket{\psi_{PS}}}{\partial B_{\alpha}\partial \bm P_{\beta}} - \frac{\partial^2 \bra{\psi_{PS}}\hat{ m}_{\alpha} \ket{\psi_{PS}}}{\partial F_{\gamma}\partial \bm P_{\beta}}  \\
     =&  -   \frac{i\hbar}{\bm M} \Bigg(\frac{\partial}{\partial B_{\alpha}} \left\langle \psi_{PS} \middle| \frac{\partial \hat{\bm\Gamma}_{\beta}}{\partial  F_{\gamma}}  \middle| \psi_{PS} \right\rangle- \frac{\partial}{\partial F_{\gamma}} \left\langle \psi_{PS} \middle| \frac{\partial \hat{\bm\Gamma}_{\beta}}{\partial  B_{\alpha}} \middle| \psi_{PS} \right\rangle\Bigg)\nonumber \\
     &+ i\hbar \frac{\partial \bm{\Pi}^{\textit{eff}}}{\partial  B_{\alpha}}   \cdot \frac{\partial^2}{\partial F_{\gamma} \partial \bm P_{\beta}}\left\langle \psi_{PS} \middle| \frac{\hat{\bm\Gamma}}{\bm M}\middle| \psi_{PS} \right\rangle\label{eq:ROA_PS2_0}\\
     =&     \frac{i\hbar}{\bm M}  \frac{\partial}{\partial F_{\gamma}} \left\langle \psi_{PS} \middle| \frac{\partial \hat{\bm\Gamma}_{\beta}}{\partial  B_{\alpha}} \middle| \psi_{PS} \right\rangle
     + i\hbar \frac{\partial \bm{\Pi}^{\textit{eff}}}{\partial  B_{\alpha}}   \cdot \frac{\partial^2}{\partial F_{\gamma} \partial \bm P_{\beta}}\left\langle \psi_{PS} \middle| \frac{\hat{\bm\Gamma}}{\bm M}\middle| \psi_{PS} \right\rangle
     \label{eq:ROA_PS2}
\end{align}
As the expression for $\hat{\bm \Gamma}$ in Eqs. \ref{eq:gammasum_mag}-\ref{eq:erf_mag} depends only on the external magnetic field, and not on the electric field, the first term on the right-hand-side of Eq. \ref{eq:ROA_PS2_0} vanishes. Note that the second term in Eq. \ref{eq:ROA_PS2} is negligible as it has a prefactor of $\bm M^{-2}$ whereas the first term has a prefactor of $\bm M^{-1}$; this extra factor of $\bm M^{-1}$ in the second term can be seen from the expression of the perturbed electronic wavefunction with respect to nuclear momentum in Eq. \ref{eq:dphidP}. The second term in Eq. \ref{eq:ROA_PS2} was also confirmed to be small for the molecule studied here and hence is ignored for the results in Fig. \ref{fig:magical_atz} below, leaving the final expression:
\begin{align}
   \frac{\partial W_{\gamma\alpha}}{\partial \bm P_{\beta}}  
     \approx &     \frac{i\hbar}{\bm M}  \frac{\partial}{\partial F_{\gamma}} \left\langle \psi_{PS} \middle| \frac{\partial \hat{\bm\Gamma}_{\beta}}{\partial  B_{\alpha}} \middle| \psi_{PS} \right\rangle
     \label{eq:ROA_PS3}
\end{align}
The programmable equations for Eq. \ref{eq:ROA_PS3} are given in the SI section S2. 

Finally, the expression in Eq. \ref{eq:ROA_PS3} requires the calculation of only the wavefunction response to the electric field, and the magnetic field derivatives of $\hat{\bm \Gamma}$ which can be seen from Eqs. \ref{eq:gammasum_mag}-\ref{eq:erf_mag} are one-electron operators. However, as we mentioned in Sec. \ref{sec:PS_W}, this expression is not independent of the gauge origin. If one is willing to sacrifice the gauge origin invariance, in Fig. \ref{fig:magical_atz}, we plotted the ROA signals calculated with Eq. \ref{eq:ROA_PS3} with a gauge origin at the center of charge of the molecule. Reasonable agreement was observed between the experiments, the perturbation theory within the BO framework, and the electronic phase space approach. 

However, note that if we set the gauge origin to be 1 \text{\AA}  off the molecule center of charge, the phase space results computed with expression in Eq. \ref{eq:ROA_PS3} can change, as shown in Fig. \ref{fig:param}. In Fig. \ref{fig:param}, we further plot origin dependence for the DO PS approach as well. Note that the polarized and depolarized CIDs do depend on the electric-dipole-electric-quadrupole polarizability $\bm A$ through the ROA invariant $\delta^2$, which require a complete basis set to remove the gauge dependence, as shown in Appendix \ref{sec_delta_G}. In practice, according to Fig. \ref{fig:param}, we find that the BO results calculated with Eq. \ref{eq:CIDm_BO} and phase space results computed with the distributed gauge origin scheme in Eq. \ref{eq:final_original_do} and Eq. \ref{eq:final_original2_do} are not sensitive to changes in the gauge origins. 

\begin{figure}[!htbp]
    \centering
    \includegraphics[width=0.9\linewidth]{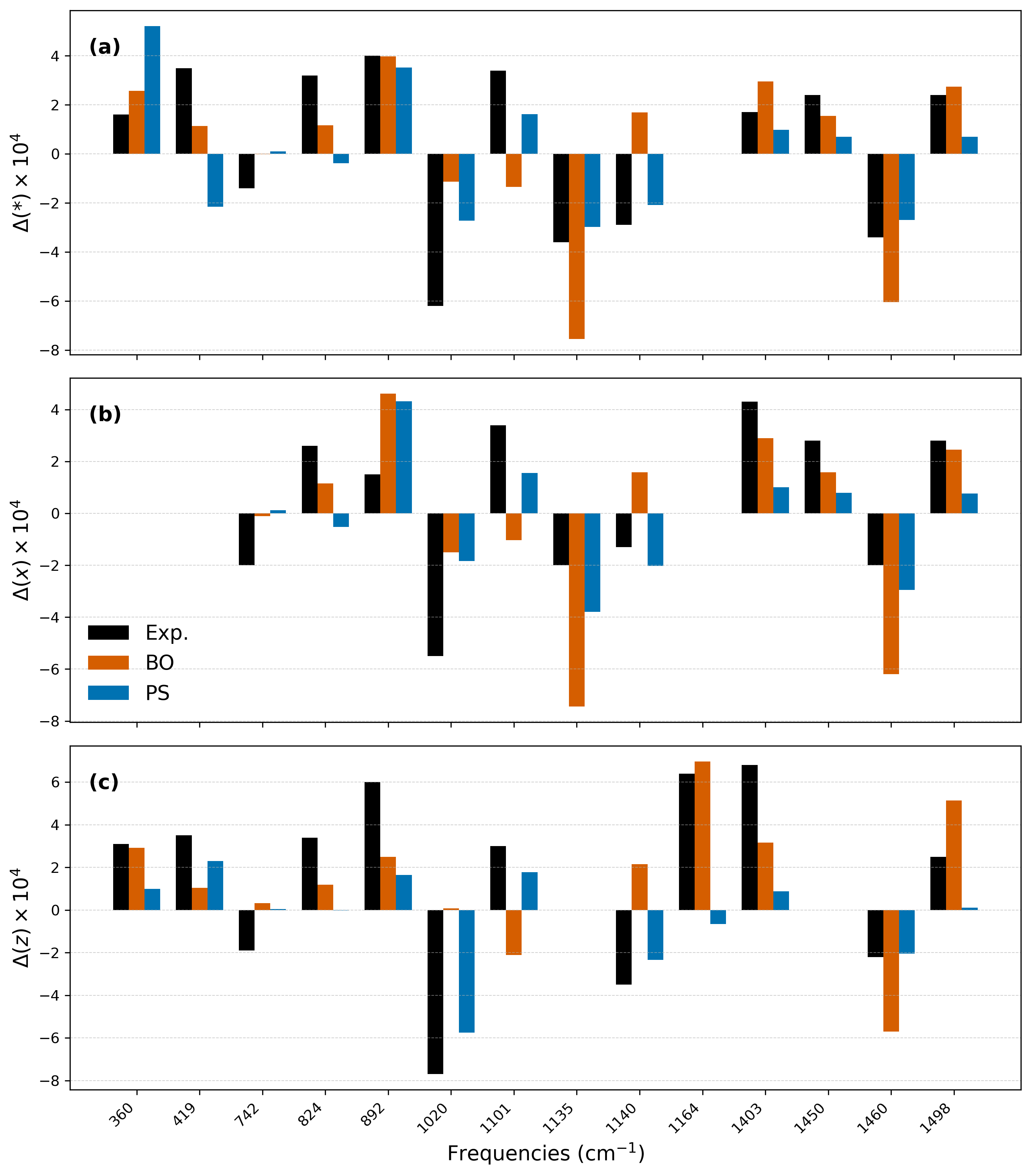}
    \caption{Calculated (a) magic-angle ($\Delta(*)$), (b) polarized ($\Delta(x)$), and depolarized CIDs ($\Delta(z)$) for (R)-methyloxirane using perturbation theory based on BO states (shown in orange) and PS theory (shown in blue, Eq. \ref{eq:ROA_PS3}, the locality parameters $\sigma=\eta=2$ Bohr and $\beta = 9$ Bohr) compared to experimental values (shown in black). The results shown here were calculated with aug-cc-pVTZ basis set. The magic-angle and polarized experimental results were reported in Ref. \citenum{bose_ab_1990}. For the depolarized experimental results, we used the experimental data from a later experiment in Ref. \citenum{polavarapu_vibrational_1993} that matched the same signs with the previous reported signals in Ref. \citenum{bose_ab_1990}.  Experimentally reported frequencies were used to label x-axis. PS results give reasonable agreement with experimental signals with the gauge center at the center of charge.  }
    \label{fig:magical_atz}
\end{figure} 

\begin{figure}[!htbp]
    \centering
    \includegraphics[width=0.9\linewidth]{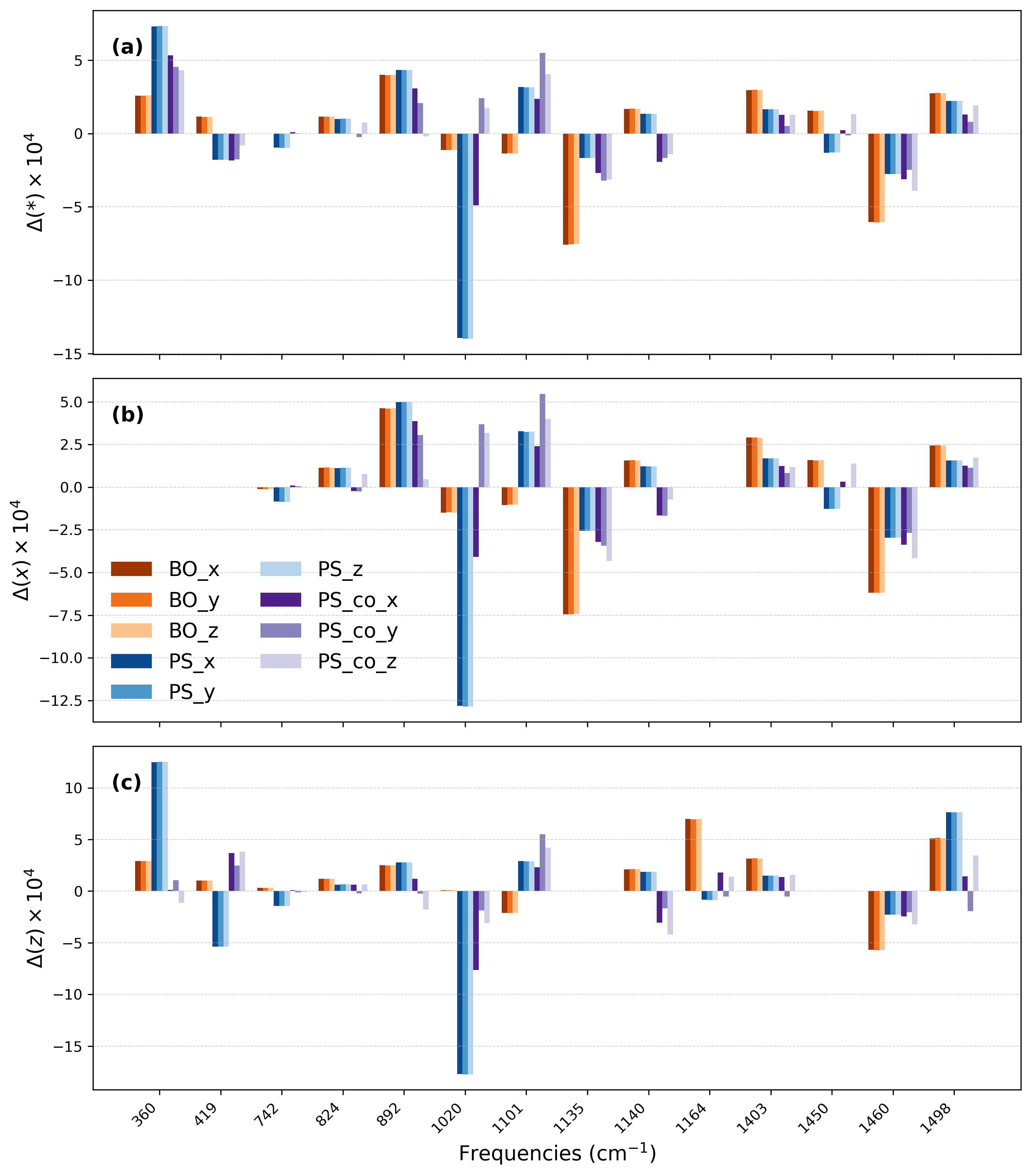}
    \caption{Calculated (a) magic-angle ($\Delta(*)$), (b) polarized ($\Delta(x)$), and depolarized CIDs ($\Delta(z)$) for (R)-methyloxirane using perturbation theory based on BO states (shown in orange), PS theory with the distributed origin scheme (shown in blue, Eqs. \ref{eq:final_original_do}-\ref{eq:final_original2_do}), PS theory with a common gauge origin (shown in purple, Eq. \ref{eq:ROA_PS3}) at different translated gauge origins. The locality parameters take the values $\sigma=\eta=2$ Bohr and $\beta = 9$ Bohr. The labels with subscript x,y,z refer to translating the molecule with 1 \AA~ along x-, y-, or z-direction.  The results shown here were calculated with aug-cc-pVTZ basis set. The CIDs computed with the BO approach and the PS approach within the distributed gauge origin scheme are not sensitive to the choice of gauge origins, despite the gauge dependence of the electric-dipole-electric-quadrupole polarizability $\bm A$, which are relevant for polarized and depolarized CID calculations. }
    \label{fig:param}
\end{figure}

\section{Supporting Information}

\subsection{Alternative Reference Frame for Electronic Pseudomomentum in Uniform Magnetic Field}
In the main text of this paper, we reviewed the recently developed phase-space approach in an uniform magnetic field $\bm B$\cite{bhati_phase-space_2025-1, bhati_phase-space_2025} and showed that the electron-nuclear dynamic coupling operator $\hat{\bm \Gamma}$ is directly related to the electronic pseudomomentum $\hat{\bm{k}}_A $ relative to some nuclear coordinates $ \tilde{\bm R}_A$.  
\begin{align}
     &\hat{\bm{k}}_A  \equiv \hat{\bm{k}}_e +  e\bm{B}\times ( \tilde{\bm R}_A-\bG)
\label{eq:pseudoAdefn_SI}
\end{align}
Moreover, rather than setting 
\begin{eqnarray}
    \tilde{\bm R}_A = \bm R_A \label{eq:X_A_SI}
\end{eqnarray} 
as we have done in the previous papers,\cite{bhati_phase-space_2025-1, bhati_phase-space_2025} we introduced local center-of-charge $\bm R_{cc}$  and center-of-mass $\bm R_{cm}$ references in Eq. 96 and Eq. 97, which we repeat here, in Eqs. \ref{eq:X_Acc} and \ref{eq:X_Acm} here. 
\begin{align}
    \bm R^{A}_{cc}  &= \frac{ \sum_B \xi_{AB} Q_B \bm{R}_B}{\sum_B \xi_{AB}Q_B } \label{eq:X_Acc_SI}\\    \bm R^A_{cm} &= \frac{ \sum_B \xi_{AB} M_B \bm{R}_B}{\sum_B \zeta_{AB}M_B }\label{eq:X_Acm_SI}
\end{align}
where $\xi_{AB}$ is a locality function
\begin{align}
\xi_{AB} =  e^{-|\bm{R}_A-\bm{R}_B|^2/\eta^2}
\label{eq:xi_SI}
\end{align}

We will now show that this change of  reference frame retains gauge invariance while conserving energy and momentum. Further, we compare the results calculated with the electronic pseudomomentum relative to each nuclei and relative to a local center-of-charge.

\subsubsection{Gauge Invariance}
We know that without the $\hat{\bm \Gamma}$-dependent term, the electronic BO Hamiltonian in a uniform magnetic field is gauge co-variant and the BO energy is gauge invariant. Mathematically, if 
\begin{eqnarray}
        \hat{H}_{BO}(\bX,\bB,\bG) \Psi(\br;\bX,\bG) =   E_{BO}(\bX,\bB,\bG) \Psi(\br;\bX,\bG)
\end{eqnarray}
then
\begin{align}
    & & \hat{H}_{BO}(\bX,\bB,\bG+ \bm{\Delta}) \Psi(\br;\bX,\bG)\exp(\frac{ie}{2\hbar}(\bB \times \bm{\Delta}) \cdot \br) = \nonumber\\
& & \; \; \; \; \; \; \; \; \; \; \; \; \; \; \; \; \; \; \; \; \; \;   E_{BO}(\bX,\bB,\bG) \Psi(\br;\bX,\bG)\exp(\frac{ie}{2\hbar}(\bB \times \bm{\Delta}) \cdot \br) 
\end{align}
Now, note that changing the reference frame of the electronic pseudomomentum does not affect the gauge covariance of the  $\hat{\bm \Gamma}$ operator because:  
\begin{align}
     \hat{\bm k}_{A}(\bm G_0+\bm\Delta,\bm B) &= \Big(\hat{\bm p} - \frac{e}{2} \bm B\times(\hat{\bm r} -\bm G_0 -\bm \Delta) + e\bm{B}\times ({\tilde{\bm R}_A}-\bG-\bm\Delta)\Big) \\
    & =\Big(\hat{\bm p}+\frac{e}{2} (\bm B\times \bm\Delta) - \frac{e}{2} \bm B\times(\hat{\bm r} -\bm G_0 -\bm \Delta)+ e\bm{B}\times ({\tilde{\bm R}_A}-\bG-\bm\Delta)\Big)   \\
    & = \hat{\bm k}_{A}(\bm G_0,\bm B)
\end{align}
Thus,  the phase-space Hamiltonian with the $\hat{\bm \Gamma}$-dependent term is still gauge co-variant and the  energy remains gauge invariant. 

\subsubsection{Translational and Rotational Invariance of Energy}
In Ref. \citenum{bhati_phase-space_2025-1}, we have previously shown that with the $\hat{\bm\Gamma}$ constraints defined in Eqs. 91 and 93 (as given in Eqs. \ref{eq:Gamma_uv2_SI} and \ref{eq:Gamma_uv4_SI}),
\begin{align}
        \label{eq:Gamma_uv2_SI}
    \Big[-i\hbar\sum_{B}\pp{}{{R}_{B\alpha}} + \hat{p}_\alpha -i \hbar \frac{\partial}{\partial G^{0}_\alpha} , \hat{ \Gamma}_{A\beta}\Big] &=0 \\
        \Big[-i\hbar\sum_{B}\left((\bm{R}_B-\bG) \times\pp{}{\bm{R}_B}\right)_{z} + \left((\hbm{r} - 
    \bG) \times \hbm{p}\right)_{z} + \hat{s}_{z}, \hat{\Gamma}_{A \delta}\Big] 
     &= i\hbar \sum_{\alpha} \epsilon_{\alpha z \delta} \hat{\Gamma}_{A \alpha}
     \label{eq:Gamma_uv4_SI}
\end{align}
the total energy is translationally and rotationally invariant, i.e.,
\begin{align}
        \sum_A \Big(\frac{\partial E_{PS}}{\partial \bX_A}\Big)_{\bm{\Pi}^{\textit{eff}}} + \Big(\frac{\partial E_{PS}}{\partial \bG}\Big)_{\bm{\Pi}^{\textit{eff}}}  = \sum_A \Big(\frac{\partial E_{PS}}{\partial \bX_A}\Big)_{\bm{\Pi}^{\textit{eff}}} &= 0\\
\sum_{A\beta\gamma}\epsilon_{z\beta\gamma}\Big(P_{A\beta} \left(\frac{\partial E_{PS}}{\partial P_{A\gamma}}\right)_\bX+ (R_{A\beta}-G^{0}_\beta) \left(\frac{\partial E_{PS}}{\partial R_{A\gamma}}\right)_\bP\Big)  &= 0
\end{align}
The local center of charge or mass in Eq. \ref{eq:X_Acc_SI} or \ref{eq:X_Acm_SI} only depends on nuclear coordinates (just like the choice of Eq. \ref{eq:X_A_SI} previously). Furthermore, translating all the nuclei shifts the reference frame same as before.
\begin{align}
   \sum_B \frac{\partial \bm R^{A}_{cc}}{\partial \bm R_B}  =\sum_B \frac{\partial \bm R^{A}_{cc} }{\partial \bm \xi_{AB}}\frac{\partial\xi_{AB}}{\partial \bm R_B} + \sum_B \delta_{BC} \frac{ \sum_C  \xi_{AC} Q_C }{\sum_C \xi_{AC}Q_C }  = 1 = \sum_B \frac{\partial \bm R^{A}}{\partial \bm R_B} \label{eq:dXcc_dX}
\end{align}
The term involves $\frac{\partial\xi_{AB}}{\partial \bm R_B}$ in Eq. \ref{eq:dXcc_dX} goes to zero, because the locality function $\xi_{AB}$ depends on the relative distance between $\bm R_A$ and $\bm R_B$, and hence does not change with translation and rotation. Therefore the new reference frame still satisfies the same constraint in Eqs. \ref{eq:Gamma_uv2} as before and hence conserves the total energy under translation. A similar statement holds for conservation under rotation. 

\subsubsection{Momentum Conservation}
In this section, we write down the expression for the $\bm q^{\textit{eff}}$ with the new reference frame for the electronic pseudomomentum for the choice of local center of charge $\tilde{\bm R}_A = \bm {R}^{A}_{cc}$ in Eq. \ref{eq:X_Acc_SI} and the corresponding conserved linear and angular momentum. We start with the phase-space energy in a uniform magnetic field $\bm B$:
\begin{eqnarray}
\label{eq:pssh_e}
    E_{\rm PS}(\bm R,\bP, \bm B) &=& \sum_A\frac{(\bm{\Pi}_{A}^{\textit{eff}})^2}{2M_A} + V_{PS}(\bX,\bP, \bm B) \\\label{eq:std_e}
    V_{PS}(\bX,\bP, \bm B) & = &  \sum_{i}\frac{{\pi}^2_i }{2} +  {V}_{eN}+ {V}_{ee} -i\hbar\sum_A \frac{\bm{\Pi}^{\textit{eff}}_A\cdot \bra{\psi_{PS}}\hat{\bm{\Gamma}}_A\ket{\psi_{PS}}}{M_{A}}- \hbar^2\sum_A \frac{\bra{\psi_{PS}}\hat{\bm \Gamma}^2_A\ket{\psi_{PS}}}{2M_A}
\end{eqnarray} 
where $V_{PS}$ is obtained by diagaonlizing the electronic phase space Hamiltonian in an external magnetic field and the corresponding eigenstate is $\psi_{PS}$. The nuclear kinetic momentum with effective nuclear charges is defined as:
\begin{eqnarray}
   \bm{\Pi}^{\textit{eff}}_A&=&\bP_A - \frac{1}{2} q^{\textit{eff}}_A(\bX)(\bB\times (\bX_A -\bG))\label{eq:pieffdefn_SI}
\end{eqnarray}
Note that here (and above and below), we always assume that $q^{\textit{eff}}_A(\bX)$ does not depend on $B$. To find an expression for the nuclear pseudomomentum, we start with a phase space equation of motion for the nuclear kinetic momentum:
\begin{eqnarray}
     \Pi^{kin,\textit{eff}}_{A\alpha} = M_A\dot{R}_{A\alpha}&=&\left(\frac{\partial E_{PS}}{\partial P_{A\alpha}}\right)_{\bm R}
    =\Pi^{\textit{eff}}_{A\alpha} -i\hbar  \left< \psi_{PS}\middle|\hat{\Gamma}_{A\alpha}\middle| \psi_{PS}\right>\label{eq:xdot}
\end{eqnarray}
This expression for the effective kinetic momentum $\bm \Pi^{kin,\textit{eff}}$ can be used to write the nuclear pseudomomentum $\bm K_{eff}$ as follows:
\begin{eqnarray}
  K^\alpha_{\textit{eff}} &\equiv&
  \sum_A \Pi^{kin,\textit{eff}}_{A\alpha} + \sum_A q^{\textit{eff}}_A(\bX)(\bB\times  (\bm{R}_A- \bm G_0))_{\alpha}
  \\
 & = & \sum_A {P_{A\alpha}}+ \frac{1}{2}\sum_A q^{\textit{eff}}_A(\bX)(\bB\times  (\bm{R}_A- \bm G_0))_{\alpha} -i\hbar  \left< \psi_{PS}\middle|\hat{\Gamma}_{A\alpha}\middle| \psi_{PS}\right>\label{eq:Kn}
\end{eqnarray}
where the $q^{\textit{eff}}_A$ is defined as (following Eq. 100):
\begin{eqnarray}
    q^{\textit{eff,cc}}_A(\bm R) =&   Q_Ae -e\sum_B\left< \psi_0 \middle|\hat{\Theta}_B(\hbr,\bm R)\middle| \psi_0 \right>\frac{  \xi_{AB} Q_A }{\sum_C \xi_{BC}Q_C }\label{eq:qeff_coc_SI}
    \label{eq:qeff}
\end{eqnarray}
Here, the $\hat{\Theta}_B$ term  partitions the space around each nuclei in such a way that the electrons in that region would also be translated or rotated with the nuclei. $\ket{\psi_0}$ here represents the ground state electronic wavefunction with $\bm B$ = 0. and $\bm P$= 0. The $\xi_{AB}$ factor  defines the pseudomomentum of electrons with respect to a localized nuclear center-of-charge around each of the atoms. 
The end result is a $\bm \Gamma$ term that is translationally invariant (which would not be possible using the standard pseudomomentum). Note that 
 Eq. \ref{eq:Kn}. 
can be rearranged
as follows:
\begin{align}
    K^\alpha_{eff} &=
  \sum_A {P_{A\alpha}}- \frac{1}{2}\sum_A q^{\textit{eff}}_A(\bX)(\bB\times  (\bm{R}_A- \bm G_0))_{\alpha} -i\hbar  \left< \psi_{PS}\middle|\hat{\Gamma}_{A\alpha}\middle| \psi_{PS}\right> +\sum_A q^{\textit{eff}}_A(\bX)(\bB\times  (\bm{R}_A- \bm G_0))_{\alpha} \\
  &= \sum_A  \Pi^{\textit{kin,eff}}_{A \alpha}  + \sum_A q^{\textit{eff}}_A(\bX)(\bB\times  (\bm{R}_A- \bm G_0))_{\alpha}  \\
  &= \sum_A  \Pi^{\textit{kin,eff}}_{A \alpha} +  Q_Ae (\bB\times  (\bm{R}_A- \bm G_0))_{\alpha} -  \sum_{B}\left< \Psi_0 \middle|\hat{\Theta}_B(\hbr,\bm R)\middle| \Psi_0 \right>\left(\frac{\zeta_{BA}Q_A  (\bB\times  (\bm{R}_A- \bm G_0))_{\alpha}}{\sum_{C}\zeta_{BC} Q_B}\right)\label{eq:Kneq}
\end{align}
where the phase space kinetic momentum is defined in Eq. \ref{eq:xdot}
and we have plugged in the definition of $\bm q^\textit{eff}$ in Eq. \ref{eq:qeff}  .
Next, we posit an electronic pseudomomentum of the form:
\begin{eqnarray}
    \bra{\psi_{PS}} k^{\alpha}_{e} \ket{\psi_{PS}} &\equiv&  \sum_{A} 
    \bra{\psi_{PS}}{k}^{A\alpha}\ket{\psi_{PS}} \\
    & = & 
      \bra{\psi_{PS}} {k}^{\alpha} \ket{\psi_{PS}} + \sum_{A}  \bra{\psi_{PS}}\hat{\Theta}_A\ket{\psi_{PS}}\frac{\sum_{B}\zeta_{AB}Q_B (\bB \times (\bX_B-\bG))_\alpha }{\sum_{B}\zeta_{AB} Q_B}\label{eq:Ke}
\end{eqnarray}
With these definitions, for a collection of nuclei with charges $q^{\textit{eff}}_A$ and electrons with charge $-e$ (we define $e > 0$), we further posit a total pseudomomentum of the form:
\begin{eqnarray}
    K^\alpha_{mol} &\equiv&
  K^\alpha_{\textit{eff}} +\bra{\psi_{PS}}
  k^{\alpha}_{e} \ket{\psi_{PS}}
\end{eqnarray}
If we add  Eq. \ref{eq:Kneq} and Eq. \ref{eq:Ke} and notice that the extra term in Eq. \ref{eq:Ke} (arising from the choice of reference for pseudomomentum) cancels approximately with the extra term in Eq. \ref{eq:Kneq} (arising from the corresponding 
change to the nuclear effective charge), it follows that
\begin{eqnarray}
    K^\alpha_{mol}
    &=&\sum_A \left( \Pi^{\textit{kin,eff}}_{A \alpha} + Q_Ae(\bB\times (\bX_A-\bG))_{\alpha} \right) + \bra{\psi_{PS}}  \hat{k}_\alpha \ket{\psi_{PS}}\label{eq:ktotmol}
\end{eqnarray}
Note that Eq. \ref{eq:ktotmol} which is the sum of the pseudomomenta of effective nuclear charges ($\bm K^{\textit{eff}}_A$) and the effective pseudomomenta of electrons ($\hat{\bm k}$) is conserved by phase space dynamics. Ref. \citenum{bhati_phase-space_2025-1} shows clearly that $\bm{K}_{mol}$ in Eq. \ref{eq:ktotmol} is conserved.  This approach makes clear that we must choose the choice of effective charge (in Eq. \ref{eq:qeff}) to match the pseudomomentum reference center (in Eq. \ref{eq:pseudoAdefn_SI}).
Likewise it can also be shown that the total angular momentum is conserved in the direction of the magnetic field. For a given magnetic field in the z-direction, the total conserved angular momentum is\cite{bhati_phase-space_2025-1}
\begin{align}
    L^{z}_{mol} &= L_{z}+\bra{\psi_{PS}} \hat{l}_{z} \ket{\psi_{PS}} \\
    L_z &= \sum_A \Bigg((\bX_A-\bG) \times \Big( \bm{\Pi}^{\textit{kin,eff}}_{A}+\frac{q^{\textit{eff}}_A(\bX)}{2}(\bB\times (\bX_A-\bG))\Big)\Bigg)_z\\
    \bra{\psi_{PS}} \hat{l}_{z} \ket{\psi_{PS}}  &= \sum_A \left<\psi_{PS}  \middle|\Big(( \hbr - \bG)\times  \hat{\Theta}_A \hat{\bm{k}}_{A}\Big)_z \middle| \psi_{PS} \right>
\end{align}

\subsubsection{ROA Circular Intensity Difference}

\begin{figure}[h!]
    \centering
    \includegraphics[width=0.9\linewidth]{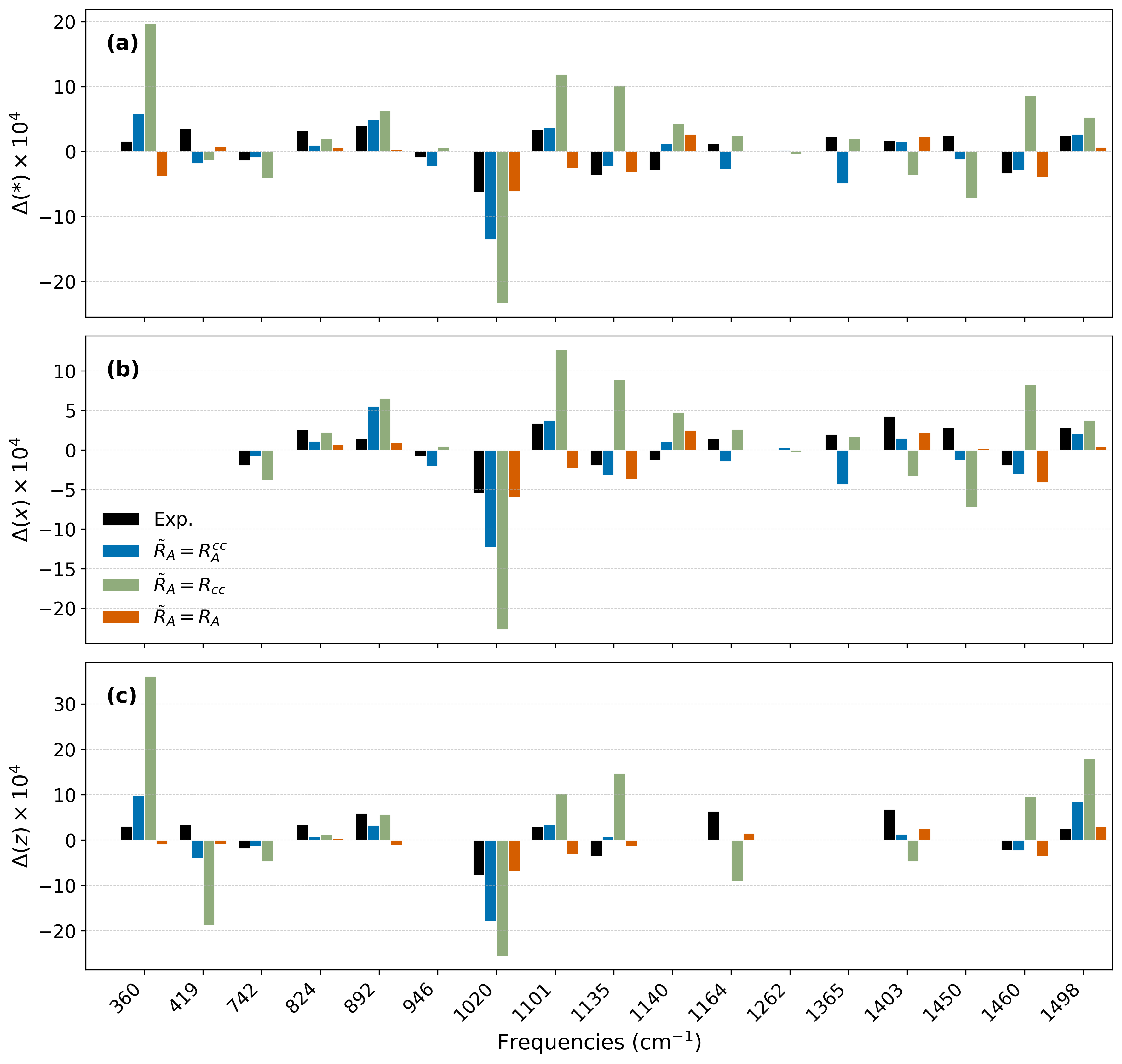}
    \caption{Calculated (a) magic-angle ($\Delta(*)$), (b) polarized ($\Delta(x)$), and depolarized CIDs ($\Delta(z)$) for (R)-methyloxirane using different reference frames for electronic pseudomomentum in the PS theory compared to experimental values (shown in black). See text for the definitions of the curves with other colors. The results shown here were calculated with aug-cc-pVTZ basis set. The magic-angle and polarized experimental results were reported in Ref. \citenum{bose_ab_1990}. For the depolarized experimental results, we used the experimental data from a later experiment in Ref. \citenum{polavarapu_vibrational_1993} that matched the same signs with the previous reported signals in Ref. \citenum{bose_ab_1990}. Experimentally reported frequencies were used to label x-axis. This data makes clear that some locality (but not strict locality) is best for establishing a reference frame.}
    \label{fig:k_psuedo}
\end{figure}

In Fig. \ref{fig:k_psuedo}, we compare the ROA circular intensity differences (CID) calculated with different reference frames for the electronic pseudomomentum to the experimental values reported in Refs. \citenum{bose_ab_1990} and \citenum{polavarapu_vibrational_1993}.  The best agreement with experimental CIDs (shown in black) was found with the calculation of the electronic pseudomomentum relative to a local center of charge (shown in blue, $\sigma=\eta = 1.5$ Bohr and $\beta = 9$ Bohr), which is also the data shown in Fig. 1 in the main paper. If the non-local limit was adopted using either the total center-of-charge (shown in green, $\sigma= 1.5$ Bohr, $\beta = 9$ Bohr, and $\eta = 20$ Bohr), the agreement with experiment becomes worse, suggesting the importance of locality. The previous choice of calculating the electronic pseudomomentum relative to the nuclear coordinates (shown in orange, $\sigma= 1.5$ Bohr and $\beta = 9$ Bohr) does not give as good match with experimental CIDs as the local center of charge, suggesting the need for some delocalization and smooth switching function. 

\subsection{Programmable equations for Gamma derivatives}
In this section, we provide the programmable equations for the atomic orbital (AO) matrix elements for the $\hat{\bm \Gamma}$ operator $ \left<\mu \middle| \hat{\bm\Gamma}\middle|\nu \right> $ and  $    \left<\mu \middle| \frac{\partial \hat{\bm\Gamma}_{\beta}}{\partial  B_{\alpha}}\middle|\nu \right> $ in a uniform external magnetic field, . The matrix elements for $\hat{\bm \Gamma}$ operator $ \left<\mu \middle| \hat{\bm\Gamma}\middle|\nu \right> $ consists of a translational coupling component and a rotational coupling component:
\begin{align}
    \left<\mu \middle| \hat{\bm\Gamma}\middle|\nu \right>  = \left<\mu \middle| \hat{\bm\Gamma}'\middle|\nu \right> + \left<\mu \middle| \hat{\bm\Gamma}''\middle|\nu \right> 
\end{align}
where 
\begin{align}
    {\hat{ \bm\Gamma}}_A' = \frac{1}{2i\hbar}\left( \hat{\Theta}_A(\hbr){\hat{\bm{ k}}_A}+{\hat{\bm{ k}}_A}\hat{\Theta}_A(\hbr)\right)\label{eq:etf_mag_SI}
\end{align}
and 
\begin{eqnarray}
     \hat{\bm{\Gamma}}_A^{''} &=& \sum_{B}  \zeta_{AB}\left(\bm R_A -\bm R^0_{B}\right)\times \left(\bm{K}_B^{-1}\frac{1}{2i\hbar}(\hbm{r}-\bm R_B)\times(\hat{\Theta}_B(\hbr) \hat{\bm{k}}_{B} + \hat{\bm{ k}}_{B}\hat{\Theta}_B(\hbr))\right)\label{eq:erf_mag_SI}
\end{eqnarray}
Recall the definition for $\hat{\bm k}_A$, 
\begin{eqnarray}
    \hat{\bm{k}}_A  &\equiv& \hat{\bm{k}}_e +  e\bm{B}\times ( \tilde{\bm R}_A-\bG) \\
    & = &\hat{\bm p} - \frac{e}{2} (\bB \times (\hbr-\bG)) +  e\bm{B}\times ( \tilde{\bm R}_A-\bG) 
\end{eqnarray}
Plugging in the definition of $\hat{\bm k}_A$ into Eq. \ref{eq:etf_mag_SI}, the AO matrix elements for $\hat{\bm \Gamma}'$ are
\begin{eqnarray}
-i\hbar\left<\mu \middle|\hat{\Gamma}'_{A\beta }\middle|\nu \right> &=& \frac{i\hbar}{2}\Big(\left< \mu \middle|\hat{\Theta}_A\middle|\nabla_{\beta} \nu \right> -\left< \nabla_{\beta} \mu \middle|\hat{\Theta}_A\middle| \nu \right> \Big)\nonumber\\&&+\frac{e}{2}\left< \mu \middle|\Big[\bB \times (\hbr-2\tilde{\bm R}_A+\bm G_0)\Big]_\beta\hat{\Theta}_A\middle|\nu \right> \label{eq:etf_nogiao}
\end{eqnarray}
The AO matrix elements for $\hat{\bm \Gamma}''$ are
\begin{align}
    -i\hbar\left<\mu \middle|\hat{\Gamma}''_{A }\middle|\nu \right> &= -i\hbar\sum_{B}  \zeta_{AB}\left(\bm R_A -\bm R^0_{B}\right)\times \left(\bm{K}_B^{-1} J^{B}_{\mu\nu}\right)
\end{align}
Here $\zeta_{AB}$ is a locality function,
  where $ \zeta_{AB} =  M_A e^{-|\bm{R}_A-\bm{R}_B|^2/\beta_{AB}^2}$. $\bm K_B $ is a moment-of-inertial like tensor, where
     $\bm{K}_B = -\sum_A\zeta_{AB}\left( (\bm{R}_A -\bm{R}^0_{B})\bm{R}_A^\top - \bm{R}_A^\top(\bm{R}_A -\bm{R}^0_{B})\mathcal{I}_3\right)$.
$\bm{R}_{B}^0$ is an averaged position, $ \bm{R}_{B}^0 =\frac{\sum_A\zeta_{AB}\bm{R}_A}{\sum_A\zeta_{AB}}$, and $\mathcal{I}_3$ is a 3 by 3 identity matrix.
The expression for matrix elements for $ J^{B}_{\mu\nu}$ is
\begin{align}
    -i\hbar J^{B \alpha}_{\mu\nu} = &\sum_{\beta\gamma}\epsilon_{\alpha\beta\gamma} \Bigg(\frac{i\hbar}{2}\Big[\left< \mu \middle|(\hbr- \bX_B)_\beta \hat{\Theta}_B\middle| \nabla_{\gamma}\nu \right> -\left< \nabla_{\gamma} \mu\middle| (\hbr- \bX_B)_\beta \hat{\Theta}_B\middle| \nu \right> \Big]\nonumber\\
    &+\frac{e}{2} \sum_{\eta\kappa}\epsilon_{\gamma\eta\kappa}\left< \mu \middle|(\hbr- \bX_B)_\beta B_{\eta} (\hbr-2\tilde{\bX}_B+\bm G_0)_{\kappa}\hat{\Theta}_B\middle| \nu \right>\Bigg) \label{eq:Juv_1}\\
    =&\frac{i\hbar}{2}\sum_{\beta\gamma}\epsilon_{\alpha\beta\gamma} \Bigg(\left< \mu \middle|(\hbr- \bX_B)_\beta \hat{\Theta}_B\middle| \nabla_{\gamma}\nu \right> -\left< \nabla_{\gamma} \mu\middle| (\hbr- \bX_B)_\beta \hat{\Theta}_B\middle| \nu \right> \Bigg) \nonumber\\
    &+\frac{e}{2} \sum_{\beta} \left< \mu \middle|(\hbr- \bX_B)_\beta B_{\alpha} (\hbr-2\tilde{\bX}_B+\bm G_0)_{\beta}\hat{\Theta}_B\middle| \nu \right> \nonumber\\
    &-\frac{e}{2} \sum_{\beta} \left< \mu \middle|(\hbr- \bX_B)_\beta B_{\beta} (\hbr-2\tilde{\bX}_B+\bm G_0)_{\alpha}\hat{\Theta}_B\middle| \nu \right>\label{eq:Juv_2}\\
      =&\frac{i\hbar}{2}\sum_{\beta\gamma}\epsilon_{\alpha\beta\gamma} \Bigg(\left< \mu \middle|(\hbr- \bX_B)_\beta \hat{\Theta}_B\middle| \nabla_{\gamma}\nu \right> -\left< \nabla_{\gamma} \mu\middle| (\hbr- \bX_B)_\beta \hat{\Theta}_B\middle| \nu \right> \Bigg) \nonumber\\
    &+\frac{e}{2} \sum_{\beta} B_{\alpha} \Big(\left< \mu \middle|\hat{r}^2_\beta \hat{\Theta}_B\middle| \nu \right> +(G^{0}_{\beta} - X_{B\beta} - 2\tilde{R}_{B\beta})\left< \mu \middle| \hat{r}_\beta\hat{\Theta}_B\middle| \nu \right> -X_{B\beta}(G^{0}_{\beta}-2\tilde{R}_{B\beta}) \left< \mu \middle| \hat{\Theta}_B\middle| \nu \right>  \Big)\nonumber\\
    &-\frac{e}{2} \sum_{\beta}  B_{\beta} \Big[\left< \mu \middle|\hat{r}_\beta\hat{r}_\alpha \hat{\Theta}_B\middle| \nu \right> +(G^{0}_{\alpha} - 2\tilde{R}_{B\alpha})\left< \mu \middle| \hat{r}_\beta\hat{\Theta}_B\middle| \nu \right> -X_{B\beta}\left< \mu \middle| \hat{r}_\alpha\hat{\Theta}_B\middle| \nu \right> \nonumber\\
    &-X_{B\beta}(G^{0}_{\alpha} -2\tilde{R}_{B\alpha})\left< \mu \middle| \hat{\Theta}_B\middle| \nu \right>  \Big]
    \label{eq:Juv} 
\end{align}
Here, to go from Eq. \ref{eq:Juv_1} to Eq. \ref{eq:Juv_2}, we evaluated the product of the Levi-Civita symbols,
\begin{align}
    \sum_\gamma \epsilon_{\alpha\beta\gamma} \epsilon_{\gamma\kappa\eta} = \delta_{\alpha\kappa}\delta_{\beta\eta} - \delta_{\alpha\eta} \delta_{\beta\kappa}
\end{align}
Now we present the programmable equations for  $    \left<\mu \middle| \frac{\partial \hat{\bm\Gamma}_{\beta}}{\partial  B_{\alpha}}\middle|\nu \right> $. We first evaluate
\begin{align}
\frac{\partial \hat{{k}}_{A\alpha}}{\partial  B_{\beta}} &= - \frac{e}{2}\sum_{\gamma}\epsilon_{\alpha\beta\gamma} (\hbr-\bG)_{\gamma}+ e \sum_{\gamma}\epsilon_{\alpha\beta\gamma} (\tilde{\bm{R}}_A-\bG)_{\gamma} \\
& = - \frac{e}{2}\sum_{\gamma}\epsilon_{\alpha\beta\gamma} \hat{r}_{\gamma}- \frac{e}{2}\sum_{\gamma}\epsilon_{\alpha\beta\gamma}G^{0}_{\gamma}+ e \sum_{\gamma}\epsilon_{\alpha\beta\gamma} \tilde{{R}}_{A\gamma} \label{eq:dkdB}
\end{align}
Taking derivative of $\left<\mu \middle|\hat{\Gamma}'_{A\beta }\middle|\nu \right>$ in Eq. \ref{eq:etf_nogiao} with respect to the external magnetic field at $\bm B=0$, we get
\begin{align}
    \frac{\partial \hat{\Gamma}'_{A\alpha}}{\partial  B_{\beta}} &= \frac{1}{i\hbar}\hat{\Theta}_A(\hbr) \frac{\partial \hat{{k}}_{A\alpha}}{\partial  B_{\beta}} \\
    -i\hbar\frac{\partial\left<\mu \middle|\hat{\Gamma}'_{A\alpha }\middle|\nu \right> }{\partial B_{\beta}}&= \frac{e}{2}\sum_{\gamma}\epsilon_{\alpha\beta\gamma}\Big(\left<\mu \middle|\hat{r}_{\gamma}\hat{\Theta}_A(\hbr)\middle|\nu \right>  + (G^{0}_{\gamma}-2\tilde{R}_{A\gamma})\left<\mu \middle|\hat{\Theta}_A(\hbr)\middle|\nu \right>\Big) \label{eq:dETFdB}
\end{align}
Similarly, taking the derivative of $\left<\mu \middle|\hat{\Gamma}''_{A\beta }\middle|\nu \right>$ in Eq. \ref{eq:Juv} with respect to the external magnetic field at $\bm B=0$, we find
\begin{align}
    -i\hbar\bra{\mu}\frac{\partial \hat{\bm\Gamma}''_{A}}{\partial \bm B}\ket{\nu} & =  -i\hbar\sum_{B}  \zeta_{AB}\left(\bm R_A -\bm R^0_{B}\right)\times \left(\bm{K}_J^{-1}(\hbm{r}-\bm R_B)\times \frac{\partial \hat{\bm\Gamma}'}{\partial \bm B}\right)\\
    & =   -i\hbar\sum_{B}  \zeta_{AB}\left(\bm R_A -\bm R^0_{B}\right)\times \left(\bm{K}_B^{-1}\frac{\partial \hat{\bm J}^{B}_{\mu\nu}}{\partial\bm B}\right) \\
     -i\hbar\frac{\partial J^{B \alpha}_{\mu\nu}}{\partial B_{\eta}} = &\frac{e}{2}\sum_{\beta\gamma\kappa}\epsilon_{\alpha\beta\gamma} \epsilon_{\gamma\eta\kappa} \Bigg(\Big[\left< \mu \middle|(\hbr- \bX_B)_\beta \hat{r}_{\kappa} \hat{\Theta}_B\middle| \nu \right> 
    + (G^{0}_{\kappa}-2\tilde{R}_{B\kappa})\left<\mu \middle|(\hbr- \bX_B)_\beta\hat{\Theta}_B(\hbr)\middle|\nu \right>\Bigg)\\
     =  &\frac{e}{2} \delta_{\alpha\eta} \Big(\sum_{\kappa} \left< \mu \middle|(\hbr- \bX_B)_\kappa \hat{r}_{\kappa} \hat{\Theta}_B\middle| \nu \right>+ (G^{0}_{\kappa}-2\tilde{R}_{B\kappa})\left<\mu \middle|(\hbr- \bX_B)_\kappa\hat{\Theta}_B(\hbr)\middle|\nu \right>\Big) \nonumber\\
     &- \frac{e}{2} \Big(\Big[\left< \mu \middle|(\hbr- \bX_B)_\eta \hat{r}_{\alpha} \hat{\Theta}_B\middle| \nu \right> 
    + (G^{0}_{\alpha}-2\tilde{R}_{B\alpha})\left<\mu \middle|(\hbr- \bX_B)_\eta\hat{\Theta}_B(\hbr)\middle|\nu \right>\Big)\label{eq:dERFdB}
\end{align} 

\subsection{ROA Circular Intensity Difference}

In Tables \ref{tab:magic}- \ref{tab:z} we report all the raw data for Fig. 1 in the main paper. The nuclear center of charge at the origin of the coordinate system for the BO approach.  Neither the BO nor distributed origin PS results are sensitive to the changes of the gauge origin $(\bm G_0)$. Note that all the PS data in Tables \ref{tab:magic}-\ref{tab:z} below use the local center of charge reference frame $ \tilde{\bm R}_A =  \bm R^{A}_{cc}$ to define the electronic pseudomomentum in Eq. \ref{eq:pseudoAdefn}.

\begin{table}[H]
\centering
\small
\setlength{\tabcolsep}{10pt} 
\caption{Magic-angle CID computed with both the conventional perturbation approach and the phase space approach using the distributed gauge origin scheme.}
\label{tab:magic}
\renewcommand{\arraystretch}{0.8}
\begin{tabular}{c c c c c }
\toprule
\multicolumn{2}{c}{\textbf{Vib. Freq. (cm$^{-1}$)}}  & \multicolumn{3}{c}{\textbf{Magic-angle CID $\Delta(*)\times 10^4$}} \\
\cmidrule(lr){1-2}\cmidrule(lr){3-5}
Hartree--Fock & Exp.\cite{bose_ab_1990} & Exp.\cite{bose_ab_1990} & BO & PS\_DO \\
\midrule
227  & 200  &      & 2.53 & -9.68 \\
397  & 360  & 1.6  & 2.58 & 5.86  \\
442  & 419  & 3.5  & 1.14 & -1.84 \\
851  & 742  & -1.4 & -0.02 & -0.91 \\
940  & 824  & 3.2  & 1.16 & 1.02 \\
986  & 892  & 4    & 3.98 & 4.89 \\
1070 & 946  &      & -1.44 & -0.90 \\
1139 & 1020 & -6.2 & -1.13 & -13.56 \\
1238 & 1101 & 3.4  & -1.35 & 3.71 \\
1267 & 1135 & -3.6 & -7.55 & -2.30 \\
1296 & 1140 & -2.9 & 1.69 & 1.21 \\
1310 & 1164 &      & 4.11 & 1.19 \\
1407 & 1262 &      & -0.46 & 0.04 \\
1526 & 1365 &      & -1.23 & 2.30 \\
1574 & 1403 & 1.7  & 2.96 & 1.48 \\
1600 & 1450 & 2.4  & 1.55 & -1.28 \\
1614 & 1460 & -3.4 & -6.05 & -2.84 \\
1672 & 1498 & 2.4  & 2.75 & 2.73 \\
3162 &      &      & 0.06 & 0.02 \\
3217 &      &      & -0.003 & 0.01 \\
3231 &      &      & -1.41 & -0.54 \\
3242 &      &      & 0.57 & 0.49 \\
3261 &      &      & 1.01 & -0.08 \\
3324 &      &      & -0.72 & -0.05 \\
\bottomrule
\end{tabular}
\end{table}

\begin{table}[H]
\centering
\small
\setlength{\tabcolsep}{10pt} 
\caption{Polarized CID computed with both the conventional perturbation approach and the phase space approach using the distributed gauge origin scheme.}
\label{tab:x}
\renewcommand{\arraystretch}{0.8}
\begin{tabular}{c c c c c }
\toprule
\multicolumn{2}{c}{\textbf{Vib. Freq. (cm$^{-1}$)}}  & \multicolumn{3}{c}{\textbf{Polarized CID $\Delta(x)\times 10^4$}} \\
\cmidrule(lr){1-2}\cmidrule(lr){3-5}
Hartree--Fock & Exp.\cite{bose_ab_1990} & Exp.\cite{bose_ab_1990} & BO & PS\_DO  \\
\midrule
227  & 200  &      & 1.92  & -7.02 \\
397  & 360  &      & 2.46  & 4.52  \\
442  & 419  &      & 1.16  & -1.49 \\
851  & 742  & -2   & -0.11 & -0.78 \\
940  & 824  & 2.6  & 1.15  & 1.11  \\
986  & 892  & 1.5  & 4.61  & 5.56  \\
1070 & 946  &      & -1.25 & -0.76 \\
1139 & 1020 & -5.5 & -1.5  & -12.26 \\
1238 & 1101 & 3.4  & -1.03 & 3.80  \\
1267 & 1135 & -2   & -7.44 & -3.16 \\
1296 & 1140 & -1.3 & 1.58  & 1.07  \\
1310 & 1164 &      & 3.54  & 1.43  \\
1407 & 1262 &      & -0.37 & 0.06  \\
1526 & 1365 &      & -1.11 & 2.00  \\
1574 & 1403 & 4.3  & 2.9   & 1.53  \\
1600 & 1450 & 2.8  & 1.58  & -1.25 \\
1614 & 1460 & -2   & -6.19 & -3.04 \\
1672 & 1498 & 2.8  & 2.45  & 2.03  \\
3162 &      &      & 0.05  & 0.02  \\
3217 &      &      & 0.03  & 0.07  \\
3231 &      &      & -1.51 & -0.64 \\
3242 &      &      & 0.5   & 0.47  \\
3261 &      &      & 0.85  & -0.08 \\
3324 &      &      & -0.83 & -0.13 \\
\bottomrule
\end{tabular}
\end{table}

\begin{table}[H]
\centering
\small
\setlength{\tabcolsep}{10pt} 
\caption{Depolarized CID computed with both the conventional perturbation approach and the phase space approach using the distributed gauge origin scheme}
\label{tab:z}
\renewcommand{\arraystretch}{0.8}
\begin{tabular}{c c c c c c}
\toprule
\multicolumn{2}{c}{\textbf{Vib. Freq. (cm$^{-1}$)}}  & \multicolumn{4}{c}{\textbf{Depolarized CID $\Delta(z)\times 10^4$}} \\
\cmidrule(lr){1-2}\cmidrule(lr){3-6}
Hartree--Fock & Exp.\cite{bose_ab_1990} & Exp.\cite{bose_ab_1990}$^{*}$  & Exp.\cite{polavarapu_vibrational_1993}$^{*}$ & BO & PS\_DO\\
\midrule
227  & 200  &      &       &  4.08 & -16.50 \\
397  & 360  & 2.5  & 3.1   &  2.92 &  9.85  \\
442  & 419  & 5    & 3.5   &  1.04 & -3.93  \\
851  & 742  & -3.1 & -1.9  &  0.32 & -1.39  \\
940  & 824  & 1.5  & 3.4   &  1.20 &  0.77  \\
986  & 892  & 4.6  & 6     &  2.49 &  3.31  \\
1070 & 946  & 8.3  &  -0.9     & -2.49 & -1.71  \\
1139 & 1020 & -5.1 & -7.7  &  0.08 & -17.91 \\
1238 & 1101 & 1.2  &  3     & -2.11 &  3.50  \\
1267 & 1135 & -3.8 &   & -7.92 &  0.76  \\
1296 & 1140 & -3.3 &  -3.5     &  2.15 &  1.76  \\
1310 & 1164 & 4    & 6.4   &  6.97 & -0.03  \\
1407 & 1262 &      &  -0.6     & -0.95 & -0.09  \\
1526 & 1365 &      &  -3.1     & -1.80 &  3.76  \\
1574 & 1403 & 3.1  & 6.8   &  3.16 &  1.34  \\
1600 & 1450 & 0.8  &       &  1.47 & -1.37  \\
1614 & 1460 & -2.7 & -2.2  & -5.70 & -2.34  \\
1672 & 1498 & 1    & 2.5   &  5.14 &  8.49  \\
3162 &      &      &       &  0.65 &  0.46  \\
3217 &      &      &       & -0.08 & -0.11  \\
3231 &      &      &       & -1.17 & -0.27  \\
3242 &      &      &       &  1.22 &  0.64  \\
3261 &      &      &       &  2.61 & -0.03  \\
3324 &      &      &       & -0.44 &  0.16  \\
\bottomrule
\end{tabular}
\end{table}
$^{*}$ Note that in all the figures, we only plotted the experimental data that give the same signs in Ref. \citenum{bose_ab_1990} and Ref. \citenum{polavarapu_vibrational_1993}. 

\subsection{Vibrational Frequencies}
In Table \ref{tab:freq}, we list the vibrational frequencies computed with Hartree-Fock (HF) and the density functional theory (DFT) with the B3LYP functional and the corresponding vibrational mode assignments. The vibrational modes obtained at the HF level generally have the same assignments as those computed with DFT/B3LYP, except for modes 10–11 and 20–22. The HF frequencies were re-ordered to align with the DFT/B3LYP assignments.

\begin{table}[h!]
\centering
\setlength{\tabcolsep}{10pt} 
\caption{Experimental and computed vibrational frequencies (cm$^{-1}$) and assignments.}
\renewcommand{\arraystretch}{0.8}
\begin{tabular}{c c c c l}
\hline
\label{tab:freq}
\textbf{Mode} &\textbf{Exp.} & \textbf{ HF} & \textbf{DFT/B3LYP} & \textbf{Mode Assignment (asterisk indicates the chiral center)} \\
\hline
1& 200  & 227  & 210  & CH\textsubscript{3} torsion \\
2&360  & 397  & 372  & C--C*--CH\textsubscript{3} bend + O--C*--CH\textsubscript{3} bend \\
3& 419  & 442  & 413  & O--C*--CH\textsubscript{3} bend + C--C*--CH\textsubscript{3} bend \\
4& 742  & 851  & 771  & C*--O str + C--O str + C*--CH\textsubscript{3} str \\
5& 824  & 940  & 843  & C*--O str + C*--CH\textsubscript{2} str \\
6& 892  & 986  & 908  & CH\textsubscript{2} twist \\
7&946  & 1070 & 972  & C--O str + C*--CH\textsubscript{3} str + CH\textsubscript{3} rock \\
8&1020 & 1139 & 1044 & CH\textsubscript{3} rock + CH\textsubscript{2} twist + C*--H bend \\
9&1101 & 1238 & 1133 & CH\textsubscript{2} rock + CH\textsubscript{3} rock \\
10&1135 & 1296 & 1156 & CH\textsubscript{2} wagging \\
11&1140 & 1267 & 1168 & CH\textsubscript{2} twisting  + C*--CH\textsubscript{3} str\\
12&1164 & 1310 & 1191 &  C*--H bend\\
13&1262 & 1407 & 1294 & C*--H bend + C*--CH\textsubscript{2} str \\
14&1365 & 1526 & 1411 & C*--H bend + CH\textsubscript{3} wagging \\
15&1403 & 1574 & 1439 & CH\textsubscript{2} sym bend +   C*--CH\textsubscript{3} str \\
16&1450 & 1600 & 1485 & CH\textsubscript{3} asym  bend \\
17&1460 & 1614 & 1498 & CH\textsubscript{3} asym  bend \\
18&1498 & 1672 & 1533 & CH\textsubscript{2} sym bend\\
19&     & 3162 & 3028 & CH\textsubscript{3} sym str\\
 20&    &3242  & 3082 & CH\textsubscript{2} sym str \\
21&     & 3217  & 3086 & CH\textsubscript{3} asym str\\
22&     &  3231 & 3088 &  C*--H str + CH\textsubscript{3} asym str \\
23&     & 3261& 3110 & C*--H str\\
24&     & 3324 & 3166 & CH\textsubscript{2} asym str\\
\hline
\end{tabular}
\end{table}

\newpage
\bibliography{main_test}

\end{document}